\documentclass[twocolumn,twocolappendix,nofootinbib,iop]{openjournal}
\usepackage{mathptmx}
\usepackage[T1]{fontenc}
\DeclareRobustCommand{\VAN}[3]{#2}
\let\VANthebibliography\thebibliography
\def\thebibliography{\DeclareRobustCommand{\VAN}[3]{##3}\VANthebibliography}
\usepackage{graphicx}
\usepackage{CJKutf8} 
\usepackage{amsmath}
\usepackage{amssymb}
\usepackage{rotating}
\usepackage{amsmath,amssymb}
\usepackage{aecompl}
\usepackage{array}
\usepackage{times}
\usepackage{url}
\usepackage{xcolor}
\usepackage{soul}
\usepackage{orcidlink}

\usepackage{booktabs}

\usepackage{tikz}
\usetikzlibrary{positioning,fit}
\tikzset{
  fc/.style = {draw, rounded corners=2pt, minimum width=30mm,
               minimum height=7mm, fill=gray!10},
  drop/.style = {draw=none, font=\footnotesize\itshape},
}
 
\usepackage[ruled,noline, noend]{algorithm2e}
\usepackage{orcidlink}
\usepackage{hyperref}
\hypersetup{colorlinks=true,citecolor=blue,linkcolor=blue,filecolor=blue,urlcolor=blue}
\DeclareMathOperator{\logit}{logit}
\newcommand{\msun}{{\rm M}_{\odot}}

\shorttitle{Radial extent of cluster impact on galaxies}
\shortauthors{Hao et al.}

\begin{document}
\journalinfo{The Open Journal of Astrophysics}

\title[Radial extent of cluster impact on galaxies]{Finding the boundary: Using galaxy membership to inform galaxy cluster extent through machine learning}
\author{Christine Hao\,\orcidlink{0009-0006-0248-1970}$^{1,2^\star}$, Stephanie O'Neil\,\orcidlink{0000-0002-7968-2088}$^{1,3,4}$, Mark Vogelsberger\,\orcidlink{0000-0001-8593-7692}$^{1}$, Vinh Tran\,\orcidlink{0009-0003-6068-6921}$^{1}$, Lamiya Mowla\,\orcidlink{0000-0002-8530-9765}$^{2}$, and Joshua S. Speagle (\textnormal{\begin{CJK*}{UTF8}{gbsn}沈佳士\ignorespacesafterend\end{CJK*}}) \,\orcidlink{0000-0003-2573-9832}$^{5,6,7,8}$}
\thanks{$^\star$ Email: \href{mailto:zhcekl@mit.edu}{zhcekl@mit.edu}\protect\\ 
\phantom{$^\star$ Email: }\href{mailto:zh102@wellesley.edu}{zh102@wellesley.edu}}
\affiliation{$^{1}$Department of Physics and Kavli Institute for Astrophysics and Space Research,
           Massachusetts Institute of Technology,
           Cambridge, MA 02139, USA\\
    $^{2}$Department of Physics and Astronomy, Wellesley College, Wellesley, MA 02481, USA\\
    $^{3}$Department of Physics \& Astronomy, University of Pennsylvania, Philadelphia, PA 19104, USA\\
    $^{4}$Department of Physics, Princeton University, Princeton, NJ 08544, USA\\ $^{5}$Department of Statistical Sciences, University of Toronto, 9th Floor, Ontario Power Building, 700 University Ave, Toronto, ON M5G 1Z5, Canada\\
    $^{6}$David A. Dunlap Department of Astronomy \& Astrophysics, University of Toronto, 50 St George Street, Toronto, ON M5S 3H4, Canada\\
    $^{7}$Dunlap Institute for Astronomy \& Astrophysics, University of Toronto, 50 St George Street, Toronto, ON M5S 3H4, Canada\\
    $^{8}$Data Sciences Institute, University of Toronto, 17th Floor, Ontario Power Building, 700 University Ave, Toronto, ON M5G 1Z5, Canada}
\vspace{0.25em}

\label{firstpage}

\begin{abstract}
The spatial extent of the environment's impact on galaxies marks a transitional region between cluster and field galaxies. We present a data‐driven method to identify this region in galaxy clusters with masses $M_{200\rm ,mean}>10^{13} \msun$ at $z = 0$. Using resolved galaxy samples from the largest simulation volume of IllustrisTNG (TNG300-1), we examine how galaxy properties vary as a function of distance to the closest cluster. We train neural networks to classify galaxies into cluster and field galaxies based on their intrinsic properties. Using this classifier, we present the first quantitative and probabilistic map of the transition region. It is represented as a broad and intrinsically scattered region near cluster outskirts, rather than a sharp physical boundary. This is the physical detection of a mixed population. In order to determine transition regions of different physical processes by training property-specific models, we categorise galaxy properties based on their underlying physics, i.e. gas, stellar, and dynamical. Changes to the dynamical properties dominate the innermost regions of the clusters of all masses. Stellar properties and gas properties, on the other hand, exhibit transitions at similar locations for low mass clusters, yet gas properties have transitions in the outermost regions for high mass clusters. These results have implications for cluster environmental studies in both simulations and observations, particularly in refining the definition of cluster boundaries while considering environmental preprocessing and how galaxies evolve under the effect of the cluster environment.
\end{abstract}

\keywords{methods: numerical -- galaxies: haloes -- galaxies: clusters: general -- galaxies: formation -- cosmology: dark matter -- cosmology: large-scale structure of universe.}

\maketitle

\section{Introduction}
In the universe, galaxy clusters are the largest structures bound by their own gravity, centered at the maxima of the universal density field. With large reservoirs of hot gas and deep gravitational potentials, the cluster environment significantly shapes the evolution and formation of nearby galaxies. Hence, galaxy clusters serve as excellent cosmic laboratories for the study of galaxy formation and evolution. To adequately use observations to study galaxy formation, understanding the spatial extent of a cluster is crucial, as it allows observers to conveniently separate cluster and field populations based on their positions relative to the closest cluster environment. Such transition is delineated by the evolution of intrinsic galaxy properties within the cluster environment a galaxy resides in. 

Observations reveal that the cluster and field environments impact galaxy formation in different ways. Particularly, cluster galaxies are older, redder, and more elliptical, while their field counterparts tend to be bluer and more spiral \citep{Dressler1980, Cooper2006, Donnari2021}. Therefore, residing in different environments can significantly alter the trajectory of a galaxy's evolution. Previous studies have investigated the temporal impact of the cluster environment, as well as the spatial extent of mergers and mass accretion in galaxy cluster outskirts \citep[e.g.,][]{ Cuesta2008, Behroozi2014,ONeil2024}. However, a systematic study has not been conducted to examine the spatial transition marked by the way intrinsic galaxy properties respond to the cluster and field environment. As a galaxy enters the cluster environment, it experiences a series of interactions with the intracluster medium of the host cluster (ICM). The truncation of a galaxy's star formation, for example, is often referred to as "quenching", during which numerous processes occur simultaneously, making it difficult to unravel. Quenching can occur both internally as mass quenching and globally as environmental quenching. The former often refers to processes within the galaxies, such as Active Galactic Nuclei feedback \citep{Croton2006} and Supernovae \citep{Larson1974}. One type of environmental quenching is ram pressure stripping, where the gas of a galaxy is stripped away by the intracluster medium, making its star formation less active compared to galaxies in the field \citep{Gunn1972,Abadi1999}. This process is governed by the galaxy's orbital parameters and intrinsic properties, as well as the physical conditions of the host cluster's environment. Other intracluster activities such as strangulation \citep{Larson1980}, the high‐speed galaxy–galaxy “harassment” \citep{Moore1996}, and tidal disruption \citep{Merritt1983} all contribute to the quenching of star formation in galaxies. According to \citet{Woo2012}, the quenching of galaxies by the halo-environment at $z = 0$ mostly depends on dark matter halo mass. The quenched fraction is reported to increase with halo mass and stellar mass. Moreover, cosmological simulations suggest that hot gas reservoirs in halos of mass $\gtrsim 10^{12}\,\msun$ drive both mass and environment quenching \citep{Gabor2014}. Observations of clusters up to $z \approx 1$ show that quenching efficiency depends jointly on galaxy stellar mass and environment, with both effects being inseparable in the densest regions \citep{PintosCastro2019}. Semi-analytic modeling further reveals that stellar mass is the primary quenching mechanism, while environmental processes modulate the quenched fraction via halo mass and local density \citep{Contini2019}. These elements involved in this process give rise to the opportunity to classify galaxies based on their cluster memberships, where galaxies that have not undergone such impacts may exhibit a different set of patterns in their intrinsic properties. As a result, the spatial extent of the cluster environment's influence can be informed by galaxy memberships. 

Observations and simulations reveal that baryon fractions within these clusters tend to stabilise early on, especially in massive ones \citep{Chiu2018, Angelinelli2023}. In observations, the baryonic fraction is measure by inferring the mass of hot gas in the intracluster medium (ICM) using observed X-ray imaging and the thermal SZ signal \citep{BIRKINSHAW1997, LaRoque2006, Allen2011}. The total mass is typically estimated using strong or weak gravitational lensing, which probes the projected mass distribution and benchmarks the calibrated cluster total mass. In simulations, all baryonic components can be directly obtained by summing up the gas and stellar contents within the cluster’s associated particles \citep{Kravtsov2012}. It is important to distinguish cluster-scale baryonic fractions and galaxy-scale baryonic properties (e.g. gas fraction, SFR, and gas/stellar metallicities), where the latter is generally easier to measure in simulations \citep{Planelles2013}. The co-evolution of galaxies and dark matter haloes can be inferred from the environmental conditions encountered by cluster members, with minimal complication from evolving baryonic contents. This provides an opportunity to investigate how galaxies transition near cluster outskirts, particularly the spatial distribution of properties that reflect environmental influence. One key insight is the shift in the probability of a galaxy being classified as a field or cluster member with distance, which can be informed by measurable characteristics \citep{Rykoff2014,Castignani2016}. Baryonic properties, such as gas fraction, colour, stellar-to-subhalo mass ratio, metallicity and star formation rate, as well as dark matter-related quantities like subhalo dark matter mass and local density, serve as valuable metrics for identifying environmental transitions and studying the transition region between field and cluster galaxies \citep[e.g.,][]{Wetzel2012, Haines2015, Oxland2024}. While it is common to infer galactic properties based on observable signatures, systematic studies that explain how the cluster environment affects galaxy membership remain limited. Such studies should ideally delineate the transition region between field galaxies and cluster galaxies, as these two categories exhibit markedly different intrinsic properties. Large-volume cosmological simulations of galaxy formation offer the opportunity to explore these transitions in detail, as they reproduce a wide range of galaxy characteristics and scaling relations representative of real-world observations. 

In particular, galaxies in the IllustrisTNG simulations exhibit a tight stellar-to-halo mass relation and realistic distributions of observable properties such as colour and magnitude, in agreement with empirical findings \citep{Pillepich2018b, Nelson2018}. Multiple galactic properties are known to evolve differently depending on whether a galaxy resides in a dense cluster or the field, motivating a more focused investigation into spatial extent of these transitions \citep{Hogg2004}. 

Previously studies have established cluster boundary definitions that inform our work. A physically motivated boundary that marks the onset of the cluster environment's dynamical structure and influence is the splashback radius ($R_{\mathrm{sp}}$). It defines the outer boundary of a halo separating infalling matter from collapsed, virialised material \citep{Diemer2014, Adhikari2014, More2015}. This radius corresponds to a sharp steepening in the halo density profile, occurring at the location where the logarithmic slope reaches a minimum. The splashback surface marks the transition between single-stream infall and multi-stream, virialised flows, thus providing a dynamical boundary for the halo \citep{Bertschinger1985, Vogelsberger2009}, which leads them to continue evolving under the influence of the cluster environment \citep{ONeil2024}. As the collision-less dark matter particles collapse gravitationally to form the dark matter haloes, they orbit the potential of the halo to form the multi-streaming region. Galaxies that enter the cluster environment have dynamics that follow the trajectory of the dark matter particles, which leads them to continue evolving under the influence of the cluster. This phenomenon shows that there exists a transition in the cluster environment that represents where galaxies enter the intracluster medium. This boundary distinguishes galaxy dynamics significantly. Therefore, dynamical galactic properties exhibit transitions across the splashback boundary. This is demonstrated by the first-infall trajectories of galaxies outside $R_{sp}$ being predominantly radial, whereas those inside mainly have more tangential and virialised orbits \citep{RadialPiccardo2024}. In our study, several dynamical properties are selected to test the extent of such effect. 

Environmental processes within the cluster environment not only impact dynamical behaviour but also significantly influence the stellar content of galaxies. As noted by \citet{Ahad2021}, environmental quenching alters the stellar mass function due to the dense surroundings and increased interactions among galaxies. Since any mechanism that perturbs or removes gas plays a pivotal role in regulating star formation, gas-related properties—i.e. gas-to-baryon mass fraction and gas metallicity—are excellent indicators of galaxy evolution \citep{Perez2023}. Thus, beyond dynamical considerations, the cluster environment's influence on baryonic properties must also be considered to fully understand the transition of galaxies near cluster outskirts. This emphasis is motivated by the fact that galaxy clusters, embedded in the Cosmic Web, act as large multi-phase gas reservoirs. The effect of the collisional gas component and gas accretion within the environment cannot be inferred purely from the dark matter density field \citep{O’Kane2024}. As galaxies fall into cluster cores, they encounter and interact with the hot intracluster medium, which drives significant transformations in their baryonic content. Consequently, the cluster boundary can also be traced via the behaviour of gas, offering a complementary probe. In particular, accreting gas at the cluster outskirts undergoes shock heating during its first infall. This process, marked by a sudden jump in the gas entropy profile, defines what is known as the \emph{cosmic accretion shock}. It arises as low-density, pristine gas collapses from void regions onto the cluster potential, forming a physical boundary delineated by the transition to hot, collapsed gas \citep{Lau2015b}. Recent studies such as \citet{Aung2021} demonstrate that the location of the accretion shock radius—identified through entropy gradients—is systematically offset from splashback-based boundaries. Specifically, the accretion shock radius tends to lie beyond the splashback radius ($R_{\rm sp}$), as well as other common definitions such as $R_{200,\mathrm{mean}}$. These findings mark the complexity of defining a universal cluster boundary. The offsets among $R_{\rm sp}$, $R_{\rm shock}$, $R_{200,\mathrm{mean}}$ and $R_{200,\mathrm{crit}}$ suggest that each boundary metric traces distinct physical processes. Therefore, one of the goals of this study is to evaluate whether these boundary definitions reliably capture the transition between cluster and field galaxies in both dark matter and baryonic contexts.
Thus, due to the multitude of environmental components and the complex nature of galaxy evolution, the boundary at which galaxies transition from the field to the cluster environment is inherently non-trivial. Given such complexity of physical processes at cluster outskirts and the multi-phase structure of the cluster environment, we propose a machine learning (ML) framework to find a probabilistic transition region between cluster and field populations around cluster outskirts. The distinct physical properties and evolutionary histories of field and cluster galaxies create an opportunity for pattern recognition through ML methods. In recent years, ML techniques have seen increasing application in data-driven cosmology and astrophysics, particularly in galaxy classification tasks \citep[e.g.][]{Farid_2023}. We employ a supervised learning approach using Deep Neural Networks (DNNs) to infer the spatial extent of the cluster environment's impact on galaxy evolution. Galaxy memberships---i.e., whether a galaxy is influenced by the cluster environment---serve as classification labels. Selected galactic intrinsic properties that are sensitive to environmental effects are used as input features for the neural network. The selection of input features plays a pivotal role in capturing the probabilistic transition across the cluster boundary. Our approach is motivated by two recent developments. First, the cluster outskirts host a kinematically mixed population of pristine infalling galaxies and processed orbiting galaxies \citep{Aung2020}. Second, the gas accretion shock, where processes like ram-pressure stripping can initiate quenching \citep[e.g.][]{Zinger2018}, often extends significantly beyond the dark matter splashback radius \citep{Aung2021}. This results in a physically consequential region where the environment is already shaping galaxy properties. We therefore need a new tool that expands beyond a hard boundary and is instead sensitive to the physical properties of galaxies themselves. Our work introduces a probabilistic boundary designed precisely to mark this transitional zone using intrinsic galaxy properties. 

In this paper, we use a data-driven framework to explore the spatial extent of the galaxy cluster environment's impact on galaxies by examining the transition between cluster and field galaxies. In particular, we illuminate the relationship between intrinsic galaxy properties and where the galaxy situates relative to the closest cluster. On top of that, we evaluate the spatial extent using statistical methods to probe the probabilistic transition between cluster and field galaxies. By applying this algorithm on various property sets, we also investigate the transition region depending on different intrinsic physical processes, i.e. stellar, gas, and dynamical. The paper is ordered as follows. In Section~\ref{sec:methods}, we describe the simulation (\ref{sec:simulations}), cluster and galaxy samples (\ref{sec:samples}), galaxy properties as a function of distance-to-closest-cluster (\ref{sec:galpropdist}), and ML model training, and evaluation (\ref{sec:model_construction}). Results are presented in Section~\ref{sec:results} for the convergence test (\ref{sec:conver_test}), the spatial extent of the cluster environment's influence determined using the fiducial six-property model (\ref{sec:spatial-extent}), the outcome of various tests we conduct (\ref{sec:scatter} and \ref{sec:prop_based}), and discussion of the mass dependence (\ref{sec:mass_trend}). Limitations and future work are outlined in Section~\ref{sec:discussion}. Finally, our findings are summarised in Section~\ref{sec:conclusions}.

\section{Methods}
\label{sec:methods}

In this paper, we find the transition region between galaxy cluster membership using a binary classification neural network model, trained on data from the IllustrisTNG (TNG) simulations. We choose a neural network model due to its high performance with large-scale problems and its compact nature once trained, which can be efficiently evaluated and analysed. In this section, we briefly describe the TNG simulations, our criteria for cluster and galaxy sampling, the structure of the machine learning model, its training and validation processes, and the evaluation. 

\subsection{Simulations}\label{sec:simulations}
The data analysed in this study originates from the largest high-resolution simulation volume of the TNG cosmological simulation, TNG300-1 \citep{Springel2017, Naiman2017, Pillepich2018b, Marinacci2017, Nelson2017}. The state-of-the-art IllustrisTNG project is a cosmological magnetohydrodynamic simulation suite built upon an updated version of the Illustris galaxy formation model and the moving mesh code AREPO \citep{Springel2010, Weinberger2020}.

In this work, we examine various intrinsic galaxy properties provided by the simulations catalogue, by using the full-physics variant with hydrodynamics to capture all baryonic properties of subhaloes. The cosmological parameters used are consistent with \citet{PlanckCollaborationXIII2016} : $\Omega_m = \Omega_{\text{dm}} + \Omega_b = 0.3089$, $\Omega_b = 0.0486$, $\Omega_\Lambda = 0.6911$, $\sigma_8 = 0.8159$, $n_s = 0.9667$, and the Hubble constant $H_0 = 100h \text{ km s}^{-1} \text{ Mpc}^{-1}$ where $h = 0.6774$.

The updated galaxy formation model in TNG is an evolution of the original Illustris project \citep{Vogelsberger2014b}, which reproduced galaxy properties spanning different environments \citep{Vogelsberger2014a}. The new galaxy formation model features a reconfigured supernova wind model \citep{Pillepich2018a}, a new radio mode active galactic nuclei (AGN) feedback scheme \citep{Weinberger2017}, and several fine-tuning adjustments that optimise numerical scheme convergence \citep{Pakmor2016}. TNG300-1, in particular, hosts $2500^3$ gas cells with a target cell mass of $1.1 \times 10^7\ \msun$ and $2500^3$ dark matter particles, each with a mass of $5.9 \times 10^7\ \msun$. The dark matter particles have a Plummer-equivalent gravitational softening length of $1.5$ kpc in physical units when $z \leq 1$ and in comoving units when $z > 1$. In contrast, the gas cells possess an adaptive comoving softening length with a minimum of $0.37$ kpc.

\subsection{Galaxy and cluster samples}\label{sec:samples}
The TNG simulation suite identifies groups using the Friends-of-Friends (FoF) algorithm \citep{Davis1985}, where particles are grouped together by their spatial distribution. The \textsc{SubFind} Algorithm \citep{Springel2001,Dolag2009} identifies subhaloes based on gravitationally bound particles. The Friends-of-Friends (FoF) and \textsc{SubFind} algorithms are ran by the TNG Collaboration, which produce the group catalogue used in this work. From this catalogue, we first select samples out of all the groups and subhaloes at redshift $z = 0$. The selection criteria is described as follows. Haloes with \(M_{200, mean}\) (at $z=0$) > \( 10^{13} \, \msun \) are considered clusters, and subhaloes with \( M_{\star} \) > \( 10^{9} \, \msun \) and  \( M_{dm} \) > \( 10^{9.5} \, \msun \) are considered galaxies. The thresholds for the galaxy definitions was motivated by the resolution limit, using subhalo mass functions, as described in \citet{Nagai2005}. We then group the cluster samples into four mass bins based on their $M_{200\rm,mean}$ at $z=0$: $10^{13 \text{--} 13.5} \msun$, $10^{13.5 \text{--} 14} \msun$, $10^{14 \text{--} 14.5} \msun$, and $10^{14.5 \text{--} 15} \msun$. These mass ranges are used for further analysis throughout the paper. They are chosen due to the lack of haloes with mass greater than $10^{15} \msun$ to form a mass bin of $10^{15-15.5} \msun$ . The mass bins were chosen to contain enough cluster samples for stacking, which prevented us from including haloes with mass greater than $10^{15} \msun$.

Even though the simulation associates all subhaloes with host groups, these host groups may not suffice to be a cluster environment. Hence, we use a sorting algorithm to match each galaxy to the nearest cluster. The algorithm is applied to all subhaloes and matches them to our selected clusters. This way, we can ensure that the selected galaxies are paired with their true closest cluster within the box, instead of only relative to the set of chosen galaxies. We calculate the point-to-point distance in a periodic box. Algorithm \ref{alg:galaxy_distances} in Appendix~\ref{apx:distance_alg} takes in cluster and subhalo positions in 3 dimensions and calculates the distance-to-closest cluster for each subhalo. The calculated distance is one dimensional, which is then normalised using the $R_{200,\rm mean}$ of the host cluster.

\subsection{Relating properties of galaxies to distance}\label{sec:galpropdist}
For this study, we select fifteen intrinsic galaxy properties at $z=0$. They are listed and defined in Table~\ref{tab:feature_transforms}. To gain insights on the spatial separation of galaxy populations, we investigate the behaviour each property exhibits as a function of the cluster centric distance normalised by $R_{200,\rm mean}$. It motivates our approach for probing the spatial extent of the cluster environment using galaxy properties 

Figure~\ref{fig:radial_trends} shows the distribution of intrinsic galaxy properties as a function of normalised cluster-centric distance $r/R_{200,\mathrm{mean}}$. Each panel is also overlaid with the median trend line shaded with the 16th – 84th percentiles. The galaxy population regime is described as the following:
\begin{figure*}
  \centering
\includegraphics[width=\linewidth]{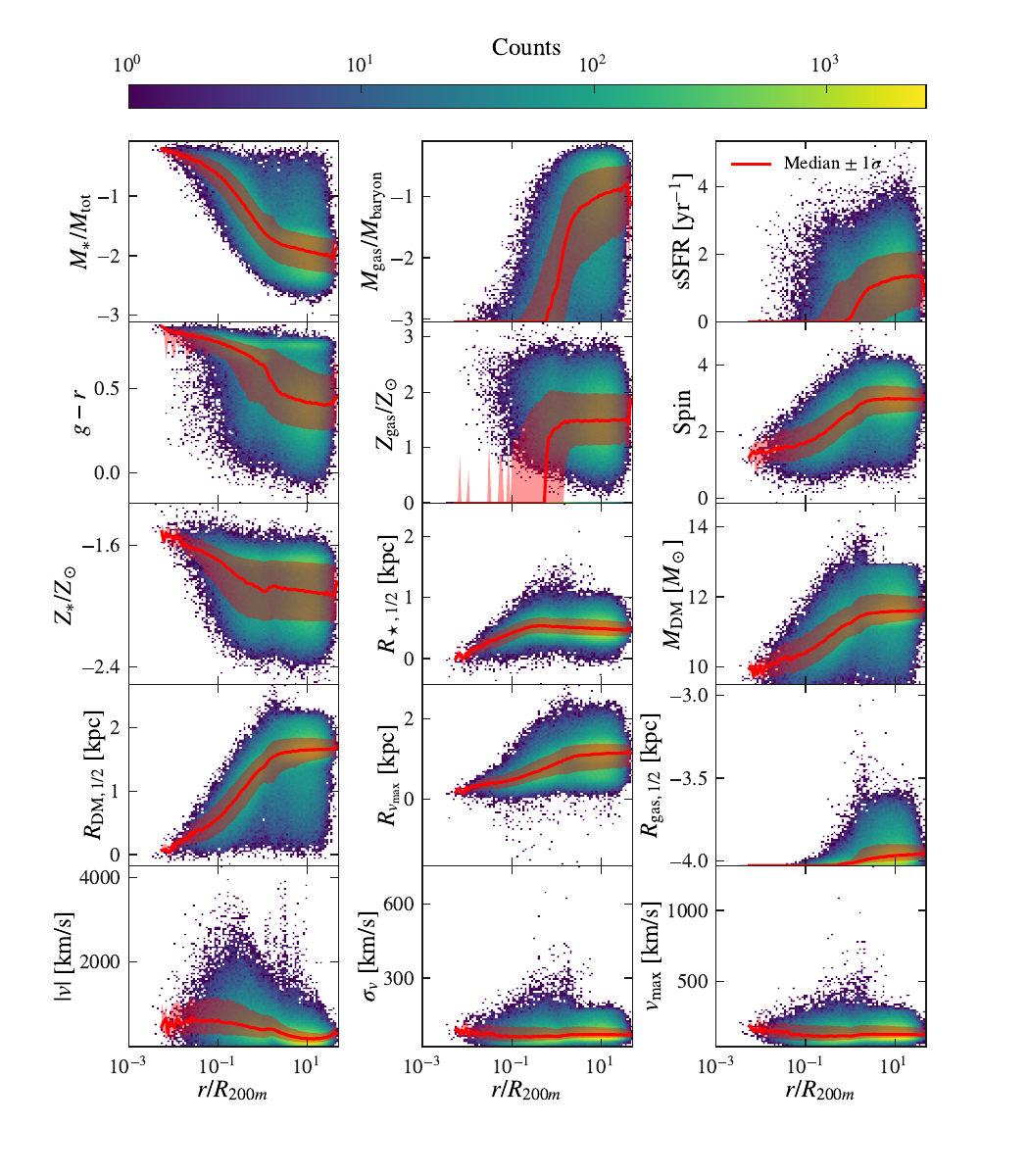}
\caption{Spatial distributions of all fifteen intrinsic galaxy properties. The colour scale indicates number density. Each panel denotes a property, as labelled on the y-axis with the corresponding units if applicable. The x-axis denotes the distance-to-closest-cluster normalised by the $R_{200,\rm mean}$ of the host. Red line denotes the median shaded by 16th-84th percentiles. The plot shows that galaxy properties correlate with galaxy-to-closest-cluster distance. The list of properties is specified in Table~\ref{tab:feature_transforms}.}
  \label{fig:radial_trends}
\end{figure*}
\begin{enumerate}
  \item \textbf{Cluster‐core population} ($R\lesssim0.5\,R_{200,\mathrm{mean}}$).  
    A compact, high‐density locus at small radii appears in several panels, such as gas‐fraction and specific star‐formation rate (sSFR), indicating a predominantly quenched, gas-stripped, early‐type cohort of cluster galaxies \citep[e.g.][]{Dressler1980, Gomez2003, Boselli2006}. The $g-r$ colour also demonstrate a narrow dispersion, which reflects the uniformity of environmental processes (e.g., ram‐pressure stripping) within the virialised core. Gradient behaviours in the median lines across several panels indicate the cluster environmental effects on galaxies strengthens as they reside closer to the core. This is consistent with recent works done in IllustrisTNG regarding quenched fractions \citep{Donnari2021}.
  
  \item \textbf{Transition population} ($1\lesssim R\lesssim3\,R_{200,\mathrm{mean}}$).  
    Between one and three times virial radii ($R_{200,\mathrm{mean}}$), the histograms broaden and often bifurcate into overlapping substructures: residual core remnants and an emerging field‐like component.  This bifurcation evidences ongoing “pre‐processing” and partial quenching as galaxies infall along large‐scale filaments. This is an instrumental part of our work.
  \item \textbf{Field‐dominated population} ($R\gtrsim 3 \,R_{200,\mathrm{mean}}$).  
    At larger radii, the distribution coalesces into a more diffused region characterised by high gas‐fractions, elevated sSFR, bluer broad‐band colours, and lower stellar‐to‐halo mass ratios. It is important to note that the galaxies between $3 \,R_{200,\mathrm{mean}}$ and $5 \,R_{200,\mathrm{mean}}$ of their host cluster do exhibit field-like properties, even though the trend-lines stabilize beyond $5 \,R_{200,\mathrm{mean}}$. These populations are indicative of isolated, late‐type systems largely unaffected by cluster‐scale environmental mechanisms \citep[e.g.][]{Sulentic2006, Boselli2006}.
\end{enumerate}

The overlaid median curves further quantify the gradual modulation of galaxy properties with cluster‐centric distance.  For instance:
\begin{itemize}
  \item \emph{Gas‐fraction and sSFR} increase sharply from the core to $R\sim2\,r_{200,\mathrm{mean}}$ before asymptoting to field values.
  \item \emph{Stellar mass and dark‐matter concentration} decline steadily toward smaller radii, reflecting enhanced stripping and mass loss in the cluster environment.
  \item \emph{Broad‐band colour and metallicity} medians shift monotonically, becoming progressively bluer at larger radii of the cluster. Stellar metallicity becomes less prominent at larger cluster radii, where gas metallicity dominates. 
\end{itemize}

Together, the juxtaposition of discrete histograms and a continuous median gradient encapsulates both the bimodal nature of cluster versus field populations and the environmental transformation pathways that operate over a continuum of cluster‐centric distances.

\subsection{Label selection}\label{sec:labelselect}
In this work, we employ Deep Neural Networks to complete the binary classification task using supervised learning. We must therefore label galaxies with their ``true'' classification (i.e. cluster or field) to construct a training set.
As part of the label selection, we express the \emph{galaxy-to-cluster} distance r in terms of a defined cluster radius $R_{x}$, where the subscript $x$ specifies the definition
adopted (\,$200\mathrm{mean}$, $200\mathrm{crit}$, or $\mathrm{sp}$\,)\footnote{$R_{\rm sp}$ is computed using the method outlined in \citet{More2015} for selected clusters.}. 

The extent of a dark matter halo is often defined using a spherical overdensity 
criterion. $R_{200,\mathrm{crit}}$ is the radius enclosing 200 times the critical density of 
the Universe ($R_{200\mathrm{c}}$), and $R_{200,\mathrm{mean}}$ the mean matter density ($R_{200,\mathrm{m}}$), where $\rho_{\mathrm{crit}}(z) = 3H^2(z)/(8\pi G)$ and 
$\rho_{\mathrm{m}}(z) = \Omega_{\mathrm{m}}(z)\,\rho_{\mathrm{crit}}(z)$.
These definitions yield systematically different halo radii, particularly at low redshift \citep{Bryan1998, More2011, Kravtsov2012}. On the other hand, there are two standard ways to compute $R_{sp}$. The first is the steepest-slope based method introduced in \citet{More2015}, which uses a spherically averaged halo density profile and takes the logarithmic slope. The other is SPARTA, outlined in \citet{Diemer2017a}, which is a particle tracking algorithm that records the first apocentres after infall of all particle orbits and takes the central value to be $R_{sp}$.

Since these boundary definitions do not necessary agree, the location where galaxies transition in and out of the cluster environment is non-trivial. Hence, we utilise these three boundary definitions as a part of a distance-based label scheme. 
We normalise the cluster-centric distance using the aforementioned radii definitions. These quantities are adopted throughout the paper when
defining cluster and field labels.
It is the primary variable that underpins our label-selection scheme. Note that we primarily use $r/R_{sp}$ during the construction and training phases of fiducial model. $r/R_{200,\mathrm{mean}}$ and $r/R_{200,\mathrm{crit}}$ are mainly employed during the post-training analysis and testing. We denote these normalisations by $r/R_x$ in general.

We label each sample that have a $r/R_x$ that falls below some threshold $(r/R_x)_{\rm cluster}$ as a cluster galaxy. Samples with an $r/R_x$ that is above our larger threshold $(r/R_x)_{\rm field}$ are labelled field galaxies. Each sample is assigned a binary label, 0 or 1, with 0 denoting a cluster galaxy and 1 denoting a field galaxy. These labelled sets are also used as a part of the convergence test discussed in depth in Section~\ref{sec:conver_test}. We vary the label–selection scheme because, even in observations, the
distinction between \emph{cluster} and \emph{field} galaxies is not
unique. Membership assignments are complicated by projection effects, redshift uncertainties, and the dynamical substructure \citep{Dressler1988, Castignani2016, Sunayama2023}. Note that the distance-based label scheme does not generate labels for all galaxies. Therefore, there are unlabelled galaxies that lie in between the lower and upper thresholds. 

In addition to the distance-based label variations, we also incorporate a labelling scheme based on the \textsc{SubLink} merger tree algorithm \citep{Rodriguez-Gomez2015}. We adopt this second method because it is how the simulation “labels” galaxies. A galaxy is considered cluster, label 0, if it has ever been associated with a host group with $M_{200,\mathrm{mean}} > 10^{13}\ \msun$ along the main progenitor branch in the merger tree. Otherwise it is a field galaxy (label 1). 

Under the distance-based label scheme, there are 38,305 galaxies labelled 0. The number of label 1 galaxies vary from 203,992 to 133,088. Under the \textsc{SubLink}-based label scheme, there are 81,258 label 0 galaxies, and 192,812 label 1 galaxies. These values are later used to compute positive weights to account for class imbalance.

\subsection{Data pre-processing: Feature preparation and train-validation-test split}\label{sec:data_prep}
During model construction and hyperparameter-tuning, we train the model on six primary properties, which are $M_\ast/M_{\rm tot}$, $M_{\rm gas}/M_{\rm baryon}$, sSFR, $g\!-\!r$, $Z_{\rm gas}$, and $Z_\star$. These quantities jointly probe stellar content and growth, gas content, magnitude, and metal-enrichment history, which are expected to evolve depending on the cluster environment \citep{Cooper2008, Wetzel2013, Boselli2020, Reeves2023}. We reapply the same architecture on different combinations of properties to test its sensitivity to various physical configurations. Table~\ref{tab:feature_transforms} lists all the properties incorporated.

\begin{table*}
  \centering
  \caption{Scaling and transformations applied to input features.}
  \label{tab:feature_transforms}
  \begin{tabular*}{\textwidth}{@{\extracolsep{\fill}} l| l l l}
    \textbf{Feature} 
    
      & \textbf{Original units} 
      & \textbf{Transformation} 
      & \textbf{Category} 
       \\
    \midrule
    Mass ratio ($M_\ast/M_{\rm tot}$)  
      & —                   & $\log_{10}(x)$ & Stellar  \\
    Gas fraction ($M_{\rm gas}/M_{\rm baryon}$)
      & —                   & $\operatorname{arcsinh}(a\,x + b)$ $a=10.56,\;b=-10.56$
                           & Gas \\
    Specific star formation rate (sSFR)                              
      & yr$^{-1}$           & $\operatorname{arcsinh}(a\,x + b)$ $a=2.22\times10^{10},\;b=0$
                           &  Gas/Stellar \\
    Colour ($g-r$)                    
      & mag                 & —              & Stellar \\
    Gas metallicity ($Z_{\rm gas}/Z_\odot$)
      & —                   & $\operatorname{arcsinh}(a\,x + b)$ $a=192.99,\;b=-0.01$
                           & Gas \\
    Spin magnitude                   
      & —                   & $\log_{10}(x)$ & Dynamical \\
    Star metallicity ($Z_{\ast}/Z_\odot$) 
      & —                   & $\log_{10}(x)$ & Stellar \\
    Star half-mass radius ($R_{\rm \star, 1/2}$)            
      & kpc                 & $\log_{10}(x)$ & Stellar  \\
    Dark-matter mass ($M_{\rm DM}$)                
      & $M_\odot$           & $\log_{10}(x)$ & — \\
    Dark-matter half-mass radius ($R_{\rm DM, 1/2}$)     
      & kpc                 & $\log_{10}(x)$ & — \\
    Velocity max radius ($R_{v_{\rm max}}$)             
      & kpc                 & $\log_{10}(x)$ & Dynamical  \\

    Gas half-mass radius ($R_{\rm gas, 1/2}$)             
      & kpc                 & $\operatorname{arcsinh}(a\,x + b)$ 
                           & Gas \\
    Velocity magnitude ($|v|$)               
      & km\,s$^{-1}$        & —              & Dynamical \\
    Velocity dispersion ($\sigma_v$)             
      & km\,s$^{-1}$        & —              & Dynamical \\
    Max velocity  ($v_{\rm max}$)                   
      & km\,s$^{-1}$        & —              & Dynamical \\
  \end{tabular*}

  \footnotesize\textbf{Notes.}  
  (1) Mass ratio is the stellar mass $M_\ast$ normalised by the total mass of the subhalo $M_{\rm{tot}}$. (2) Gas fraction is defined using the ratio between the mass of gas $M_{\rm gas
  }$ and the mass of baryons $M_{\rm baryons}$. (3) Specific star formation rate is the SFR in the gas cells divided by stellar mass $M_\ast$. (4) Colour is the $g-r$ magnitude. (5) Gas metallicity is the mass-weighted average metallicity of gas cells $Z_{\rm gas}/Z_\odot$. (6) Spin magnitude is the $l-2$ norm of the subhalo's total spin in the $x$, $y$, and $z$-axis. (7) Star metallicity is the mass-weighted average metallicity of star particles $Z_{\ast}/Z_\odot$. (8) The star half-mass radius $R_{\rm \star, 1/2}$ is the comoving radius containing half of the subhalo's stellar mass. (9) Dark-matter mass $M_{\rm DM}$ is defined by the subhalo mass type 1, summing the mass of all dark matter particles bound to the subhalo. (10) Dark-matter half-mass radius $R_{\rm DM, 1/2}$ is the comoving radius containing half of the subhalo's dark matter mass. (11) Velocity max radius $R_{v_{\rm max}}$ defines the location where $v_{\rm max}$ is achieved. (12) The gas half-mass radius $R_{\rm gas, 1/2}$ is comoving radius containing half of the subhalo's gas mass. (13) Velocity magnitude is $|v|$. (14) Velocity dispersion $\sigma_v$ is the one-dimensional velocity dispersion of all particles and cells associated with the subhalo. (15) The max velocity $v_{\rm max}$ is the maximum value of the spherically-averaged rotation curve. Note that sSFR is taken into account as both gas and stellar categories. $\log_{10}$ denotes base-10 logarithm.  
  For arcsinh transforms, parameters $(a,b)$ were chosen to symmetrise each feature’s distribution using the method discuss in Section~\ref{sec:data_prep}.
\end{table*}

To symmetrise heavy‐tailed feature spatial distributions, we apply the parametrised inverse hyperbolic sine transformation (hereafter referred to as the ``arcsinh transform'') to several features vector \(y\):
\begin{equation}
  \tilde y \;=\; \operatorname{arcsinh}\bigl(a\,y + b\bigr)\,,
\end{equation}
where the scale $a$ and shift $b$ are set from each individual transformation. Details of the arcsinh transformations are provided in Appendix~\ref{apx:arcsinh_transform}. The goal here is to symmetrise each property's spatial distribution. This reduces its dynamical range, throughout which the data is distributed evenly. On top of the arcsinh transformation, several properties undergo the logarithmic transformation --- including spin magnitude, $Z_{\star}/Z_{\odot}$, $R_{\star, \rm 1/2}$, $M_{\rm DM}$, $R_{\rm DM, 1/2}$, and $R_{\rm max}$. Properties that do not need a transformation to have a desired symmetrical distribution are $g-r$, $|v|$, $\sigma_v$, and $v_{\rm max}$. The details of all transformations and relevant parameters (if applicable) are discussed in Table~\ref{tab:feature_transforms}.

Once the feature matrix and binary labels are assembled, the sample is then partitioned into training, validation, and test sets. The latter consulted only after all hyper-parameters have been fixed. Galaxies that reside in the same cluster environment are impacted similarly, so their properties may share patterns. Sorting members of the same cluster into different folds would allow the model to ``cheat'' by recognizing environmental signatures that are present in both training and validation data. To avoid environmental contamination, we ensure that every cluster’s galaxies are assigned en bloc to either the training set, the validation set, or the test set and never to more than one partition. This strategy is also adopted in many cluster environment studies using machine learning \citep[e.g.][]{sadikov2025}. We split the clusters by the following ratios: $70\%$ for training, $15\%$ for validation, and $15\%$ testing. The model is tuned on the cluster-exclusive validation set before the test step. 

\subsection{Model construction}\label{sec:model_construction}
\label{sec:construction}

The Deep Neural Network employed in this study is constructed and trained to predict whether a galaxy resides in the cluster or the field environment. Accordingly, we design a fully connected feed-forward neural network that maps six galaxy properties to a single probability of cluster membership. In later steps, all fifteen properties mentioned previously are involved in training using different feature combinations. We discuss model construction and training in the context of using the six primary properties outlined in Section~\ref{sec:data_prep}. For each galaxy we form the feature vector
\(\mathbf{x}
   =(x_1,\ldots,x_6)^\top\in\mathbb{R}^6\),
where
\begin{equation}
\begin{aligned}
\mathbf{x}
=
\left(
  \frac{M_\star}{M_{\rm sub}},\;
  \frac{M_{\rm gas}}{M_\star+M_{\rm gas}},\;
  \mathrm{sSFR},\;
  g\!-\!r,\;
  Z_{\rm gas},\;
  Z_\star
\right)
\end{aligned}
\end{equation}
Each feature is linearly rescaled using a Min-Max transform to the interval $[-1, 1]$. This scaler is fitted only on the training split to ensure numerical stability and prevent validation or test set information from leaking into the training stage. During validation, test, and the final evaluation, galaxy samples are transformed using the scaler fitted on the training set. 

The network is a four–layer feed-forward Multilayer Perceptron (MLP) (Figure~\ref{fig:mlp_architecture}); all architectural and
optimization hyper-parameters are summarised in Table~\ref{tab:hyper_and_arch}.
Regularisation is applied on three complementary fronts:

\begin{enumerate}
\item \textbf{Dropout.}  
      After every hidden layer we drop activations with probability
      \(p_{\mathrm{drop}} = 0.10\), mitigating co-adaptation and encouraging the network to form distributed representations \citep{Srivastava2014}.

\item \textbf{Weight decay.}  
      The \textsc{AdamW} optimiser introduced in \citet{Loshchilov2017} imposes an $L_{2}$ penalty of
      \(10^{-4}\) on all weights, acting as a ridge prior that discourages overly large parameters.

\item \textbf{Batch normalization.}  
      Each hidden layer is preceded by batch normalization, which keeps the internal covariances stable and provides an additional smoothing effect 
    \citep{Ioffe2015}.
\end{enumerate}
Cosine-annealing is applied as the learning rate scheduler with half-period of $T_{\max}=200$ epochs and the floor as $\eta_{\min}=9{\times}10^{-6}$ \citep{Loshchilov2016}; early stopping with a patience of five epochs halts training once the validation loss ceases to improve \citep{Prechelt2012}.
All hyper-parameters were refined manually on the validation set, guided by performance metrics. No automated search was attempted.

During forward propagation, for each galaxy, we form the input vector $\mathbf{x}\in\mathbb{R}^{6}$ comprised of
the stellar-to-subhalo mass ratio, the gas mass fraction, the sSFR, the $g\!-\!r$ colour, the gas metallicity $Z_{\rm gas}/{\rm Z}_{\odot}$, and the stellar metallicity $Z_\star/{\rm Z}_{\odot}$. It then outputs a logit $\hat y$, which is described as
\begin{equation}
\label{eq:mlp}
\begin{aligned}
z &= f_\theta(\mathbf{x})
   = \mathbf{w}_4^{\!\top}\,
     \sigma\!\Bigl(
       \mathbf{W}_3\,\sigma\!\bigl(
         \mathbf{W}_2\,\sigma\!\bigl(
           \mathbf{W}_1\mathbf{x}+\mathbf{b}_1\bigr)+\mathbf{b}_2\bigr)
       +\mathbf{b}_3\bigr)
     + b_4,\\[4pt]
\hat y &= \sigma_{\rm sigmoid}(z),
\end{aligned}
\end{equation}

\noindent
where \(\sigma\) is the ReLU, \(\hat y\in[0,1]\) is used only for metrics, and
\(\theta=\{\mathbf{W}_\ell,\mathbf{b}_\ell\}_{\ell=1}^{4}\) describes the parameters\footnote{Here $\mathbf{W}_\ell$ denotes the weights and $\mathbf{b}_\ell$ the biases.}. For back propagation, training minimises the weighted binary-cross entropy
\begin{equation}
\mathcal{L}(\theta)=
-\frac1N\sum_{i=1}^{N}\!
\Bigl[
  y_i\log\sigma(z_i)
  + w\,(1-y_i)\log\!\bigl(1-\sigma(z_i)\bigr)
\Bigr],
\end{equation}
where $y_i \in \{0, 1\}$ is the true class label ($0$ for cluster galaxies and $1$ for field galaxies), $z_i \in \mathbb{R}$ is the raw output logit after the forward pass, and $\sigma_{\rm sigmoid}(z_i) \in (0, 1)$ is the predicted probability. The positive class weight is \(w = N_{\text{neg}}/N_{\text{pos}}\). It compensates for class imbalance.

\begin{figure}
\vspace{1cm}
\centering
\begin{tikzpicture}[
  node distance = 8mm and 11mm,
  every node/.style = {font=\small},
]

\node[fc] (x)     {Input $\mathbf{x}\,(6)$};

\node[fc, below=of x]    (l1) {Linear 6$\to$48 + BN + ReLU};
\node[drop, below=2pt of l1] (d1) {$p=0.10$};

\node[fc, below=4mm of d1] (l2) {Linear 48$\to$24 + BN + ReLU};
\node[drop, below=2pt of l2] (d2) {$p=0.10$};

\node[fc, below=4mm of d2] (l3) {Linear 24$\to$12 + BN + ReLU};
\node[drop, below=2pt of l3] (d3) {$p=0.10$};

\node[fc, below=4mm of d3] (l4) {Linear 12$\to$1 (logit)};

\node[fc, below=of l4]     (sig) {Sigmoid};
\node[fc, below=of sig]    (y)   {Output $\hat y$};

\draw[->] (x)   -- (l1);
\draw[->] (l1)  -- (d1);
\draw[->] (d1)  -- (l2);
\draw[->] (l2)  -- (d2);
\draw[->] (d2)  -- (l3);
\draw[->] (l3)  -- (d3);
\draw[->] (d3)  -- (l4);
\draw[->] (l4)  -- (sig);
\draw[->] (sig) -- (y);

\end{tikzpicture}
\caption{Bird’s-eye network architecture described in Section \ref{sec:construction}.  BN refers to BatchNorm and \emph{p} is the dropout percentage.  All activations are ReLU except the output sigmoid, applied only
at inference for metric computation.}
\label{fig:mlp_architecture}
\end{figure}
\begin{table*}
  \centering
  \caption{MLP hyperparameters and detailed layer-by-layer parameter counts.}
  \label{tab:hyper_and_arch}
  \begin{minipage}[t]{0.48\textwidth}
    \centering
    \begin{tabular}{@{}lll@{}}
      \toprule
      \multicolumn{3}{c}{\textbf{Hyper-parameters}} \\
      \midrule
      Layer sizes  & $6\!\to\!48\!\to\!24\!\to\!12\!\to\!1$ & \\
      Dropout      & $p=0.10$ (after each hidden layer)     & \\
      Optimiser    & AdamW ($\eta_0=5\times10^{-5}$, $w_d=10^{-4}$) & \\
      Scheduler    & CosineAnnealing ($T_{\max}=200$, $\eta_{\min}=9\times10^{-6}$) & \\
      Batch size   & 256 & \\
      \midrule
      Positive Weights ($R_{sp}$-based)&
      \multicolumn{2}{l}{%
        \begin{tabular}[t]{@{}l@{}}
          $0.5/1.0:$ 0.22 \\
          $0.5/2.0:$ 0.25 \\
          $0.5/3.0:$ 0.29 \\
          $0.5/4.0:$ 0.34 \\
          $0.5/5.0:$ 0.39 \\
          $1.0/1.0:$ 0.35 \\
          \textsc{SubLink}: 0.41 \\
        \end{tabular}
      } \\
      \bottomrule
    \end{tabular}
  \end{minipage}
  \hfill
  \begin{minipage}[t]{0.48\textwidth}
    \centering
    \begin{tabular}{@{}l l r@{}}
      \toprule
      \multicolumn{3}{c}{\textbf{Architecture (trainable number of parameters)}} \\
      \midrule
      Layer   & Type              & Parameters \\
      \midrule
      layer1  & Linear            &   336  \\
      bn1     & BatchNorm1d       &    96  \\
      layer2  & Linear            & 1\,200 \\
      bn2     & BatchNorm1d       &    48  \\
      layer3  & Linear            &   300  \\
      bn3     & BatchNorm1d       &    24  \\
      layer4  & Linear            &    13  \\
      loss\_fn & BCEWithLogitsLoss &     0  \\
      \bottomrule
    \end{tabular}
  \end{minipage}
\end{table*}

\subsection{Model evaluation}\label{sec:eval}
Once training is complete, the model is evaluated through two distinct inferences passes: the test set and the full dataset. Both passes are evaluated using the best checkpoint model that achieves the lowest validation loss, which is also often the last checkpoint due to early stopping. The network processes the labelled test set to produce performance metrics such as the Area Under the Receiver Operating Characteristic curve (AUROC) and Area Under the Precision Recall curve (PR-AUC). These metrics further confirm the accuracy and robustness of the trained model. Details are discussed in Appendix~\ref{apx:performance_metrics}.

For the second inference pass, the full dataset, which contains galaxies that are unlabelled, is evaluated using the same checkpoint. Similar to the training, validation, and test sets, the full dataset contains six features for each galaxy. The resulting logits on the full dataset are then stored for further analysis. The results are organised by the associated clusters in order to probe their environmental effects. Therefore, each galaxy maps to a single logit, which is then stored under the home cluster of the galaxy. The procedure is outlined here:

\begin{enumerate}
\setlength{\itemsep}{4pt}

\item \textbf{Scaling and evaluation.}  
All galaxy features are rescaled using the Min–Max transformation fitted on the training set.  
The rescaled features are then passed through the best-performing classifier in evaluation mode, producing both raw outputs (logits) and corresponding probabilities obtained through the sigmoid activation.

\item \textbf{Recorded quantities.}  
Prior to the evaluation pass, three values are retained for each galaxy \(i\),:
\begin{align}
  c_i &= \text{nearest cluster identifier}, \label{eq:cluster_id}\\
  R_{i} &= \frac{R_{\mathrm{gal},i}}{R_{200\mathrm{m}}}, \label{eq:r200m}\\
  \mathbf{x}_i &\in \mathbb{R}^6 \quad \text{(feature vector)}. \label{eq:features}
\end{align}

\item \textbf{Cluster-wise grouping.}  
Galaxies are then aggregated according to their nearest FoF halo, indexed by \(h=0,1,\dots,h_{\max}\), which selects all galaxies assigned to cluster \(h\). Let $n_h$ denote the index list of galaxies belonging to cluster $h$.
For each cluster, we record a collection of its member galaxies' logarithmic normalised radii $\log_{10}(R_{i})$ and their feature vector $x_i$.
These features are again rescaled and evaluated by the classifier to obtain
\begin{align}
  \mathbf{z}_h &= f(\widetilde{\mathbf{X}}_h), \label{eq:logits}\\
  \boldsymbol{\sigma}_h &= \sigma(\mathbf{z}_h) \in (0,1)^{n_h}, \label{eq:probs}
\end{align}
where \(\widetilde{\mathbf{X}}_h\) denotes the scaled feature matrix, \(\mathbf{z}_h\) the logits, and \(\boldsymbol{\sigma}_h\) the predicted probabilities.

\end{enumerate}

\subsection{Finding the transition region}\label{sec:zerofind}

After model evaluation, we use the prediction logits and galaxy-to-cluster distance for further analysis.
For each cluster we then construct two vectors
\begin{align}
  \mathbf z_h = (\,z_g\,)_{g\in n_h} \\
  \boldsymbol{\ell}_h = \bigl(\log_{10}R_g\bigr)_{g\in n_h},
\end{align}
where $z_g$ is the raw logit and $\boldsymbol{\ell}_h$ the logged galaxy-to-cluster distance (normalised by $R_{200,\rm mean}$). As previously stated, clusters are grouped into four logarithmic mass bins.  We aim to assign each cluster mass bin a single \emph{zero–crossing radius}, which we compute in parallel by two complementary schemes: taking the cluster-to-cluster median of individual cluster zero points and finding zero points on stacked profiles, written as $r^{\rm med}$ and $r^{\rm stack}$ respectively. All zero-point computations were performed separately within each mass range to reveal mass‐dependence.
We describe these schemes in more detail below.

\begin{figure}
    \centering
    \includegraphics[width=\linewidth]{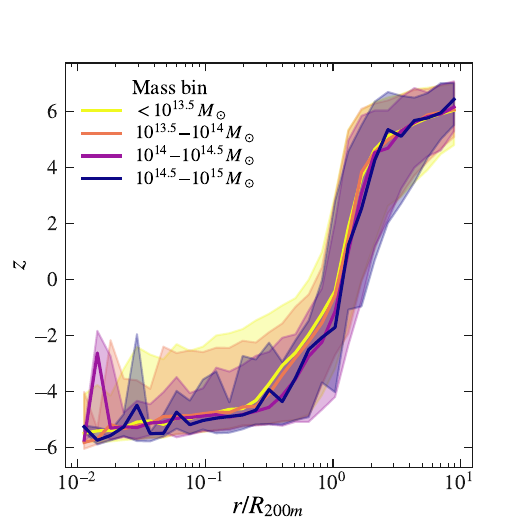}
    \caption{Per-mass bin median probability profile of the fiducial model using distance-based labels ($0.5/5.0$). The shaded region represents the 16-84 percentile scattering around the median. The y-axis is the raw logits (z) predicted by the model for each galaxy sample. Retaining the raw logits for the y-axis allows the flattening at two ends to be demonstrated more clearly. The $x$-axis denotes distance normalised by $R_{200,\rm mean}$. Logits are first collected to make the per-cluster profile based on the host cluster of each galaxy, which is then stacked into mass bins for the final profile. This clearly demonstrates the probabilistic relationship between the normalised galaxy-to-cluster distance and galaxy membership (cluster or field).}
    \label{fig:profiles-combined}
\end{figure}

\textit{Stacked–profile.}  
   We first align cluster profiles in each mass bin on a common grid by radial binning $log(r)$, and then compute a stacked probability profile for each cluster mass bin. The result for each mass bin is denoted as $P(r)^{i}_{\rm stack}$, where $i$ indicates the mass range. The stacked profile of model $0.5/5.0$ is shown in Figure~\ref{fig:profiles-combined}.

\textit{Cluster–to–cluster median.} 
We then perform cluster-to-cluster zero point analysis to find the median zero point in each mass bin. Since each mass bin has a corresponding zero point and 1$\sigma$, we store the median mass in each mass bin to illuminate the mass dependence of the transition region.
\noindent In subsequent analysis, we compute both $r_0$ and $P(r)_{\rm stack}$. 
This dual approach allows us to (i) preserve halo‐to‐halo scatter via $r_0$ and (ii) characterise ensemble behaviour via the stacked profile. 

\section{Results}
\label{sec:results}
In this section, we discuss the transition region determined based on our fiducial model and inform a choice of the cluster boundary using galaxy memberships. 
\subsection{Convergence test: Detailed discussion}\label{sec:conver_test}

A convergence test is performed by varying the upper distance ratio threshold of field galaxy labels, using the same model construction as in Section~\ref{sec:model_construction}. The upper threshold is incremented by 0.1, creating 41 labelled sets in total. We apply the convergence test on all feature combinations, where the model is shown to converge regardless of the features we choose.

We now evaluate and discuss the outcome of the convergence test. This verifies our model’s predicted cluster transition radius are robust to the different choices of distance‐based labelling. The number of labelled samples also varies based on the labels. For each subset of trained models, we fix the network architecture, optimiser settings (learning rate schedule, weight decay, batch size), and random seed governing both weight initialization and data shuffling. We adopt the same train/validation/test split for all experiments, with early‐stopping triggered by the plateau in validation loss to avoid overfitting. Each training run yields a predicted transition radius, defined as the radial location where the median cluster-to-cluster zero‐crossing occurs for every mass bin, together with the 16th and 84th percentiles to represent the inherent scatter. 
The outcome demonstrates that in each mass bin, the prediction region (median and 16th and 84th percentile errors) does not increase as we increase the upper threshold. This shows that the model is successful at determining a spatial region of separation among the two memberships, where the difference between the predicted transition regions decreases as the upper threshold goes beyond $4$. This is also consistent regardless of the feature set the model is trained on. 

\begin{figure*}
    \centering
    \includegraphics[width=7.0in]{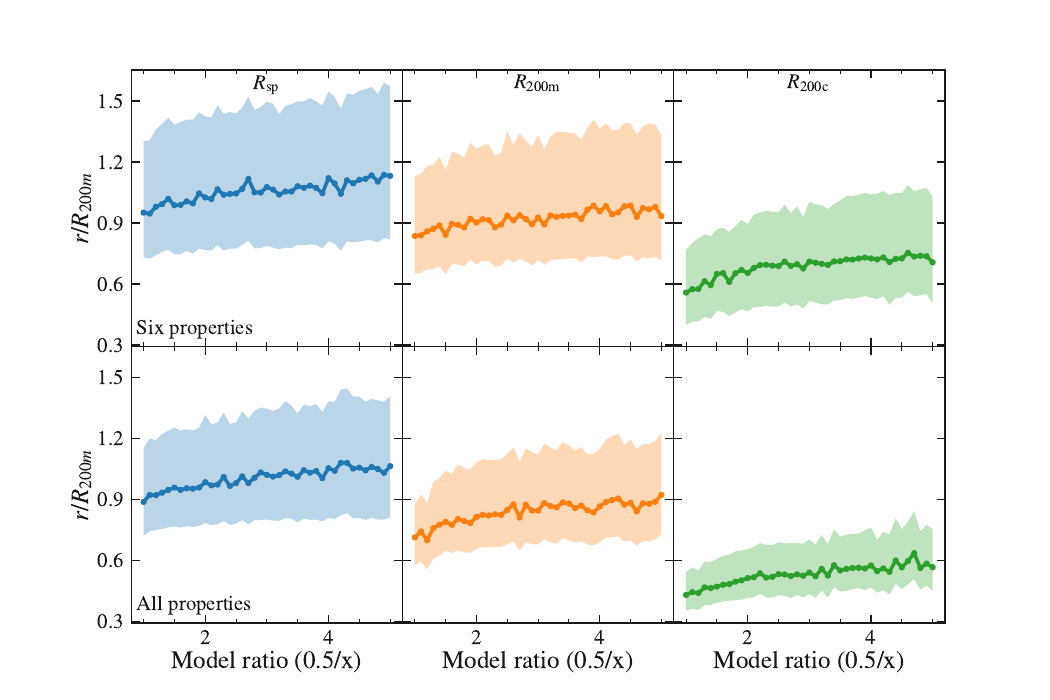}
    \caption{Convergence test curve for the fiducial model architecture using varied label upper ratio (x-axis) and the predicted transition region. Each point represents the zero point and 16th-84th percentile error in the mass bin $10^{14.0} \le M_{200,\mathrm{mean}}/M_\odot < 10^{14.5}$ predicted by each model. The upper panels denote models trained using six primary properties, while models in the lower panels are trained on all properties. The first column consists of models trained using $R_{sp}$ labels, the second $R_{200,\rm mean}$, and the third $R_{200,\rm crit}$. This shows that even when we vary the set of unlabelled samples in between the lower and upper bound, the model's prediction stays within a reasonable range and does not increase with the gap. This is consistent across different radii definitions. However, the intrinsic scatter is best represented by the $R_{sp}$-labelled models as it is the dynamically motivated definition. The $R_{200,\rm mean}$ and $R_{200,\rm crit}$ panels exhibit more fluctuation.}
    \label{fig:conver-sixprop}
\end{figure*}

Figure~\ref{fig:conver-sixprop} shows the median transition radii for a single mass bin $10^{14.0} \le M_{200,\mathrm{mean}}/M_\odot < 10^{14.5}$, plotted as a function of the incremental upper distance threshold for field galaxies. Each column denotes a different radii definition used for labelling ($r/R_{sp}$, $r/R_{200,\rm mean}$, and $r/R_{200, \rm crit}$ respectively). The same trend is observed across the other mass bins. The top row represents models trained using the six primary properties, and the bottom uses all fifteen properties. Even when the threshold increases more than five times, the shifts in the inferred radius never exceed the width of the 16–84\% bands—typically $\Delta (r/R_{200m})\lesssim0.5$. This demonstrates that our model’s prediction of the transition region is not an artifact of a particular distance-based label threshold: once a split of cluster-like and field-like galaxies are established through the labelled sets, the transition region becomes a persistent identity. Although models labelled using all three radii definitions exhibit  convergence behaviour, the $R_{sp}$-based models have the most stable trend, while successfully demonstrating the expected intrinsic scattering. Finally, the same convergence behaviour is observed across different feature subsets (e.g.\ six‐property, all‐property, gas‐only, stellar‐only, and dynamics only).  These feature combinations are added to test whether different property combinations exhibit changes at different locations.
\subsection{Spatial extent of the cluster environment's influence}
\label{sec:spatial-extent}
\begin{figure*}
    \centering
\includegraphics[width=\linewidth]{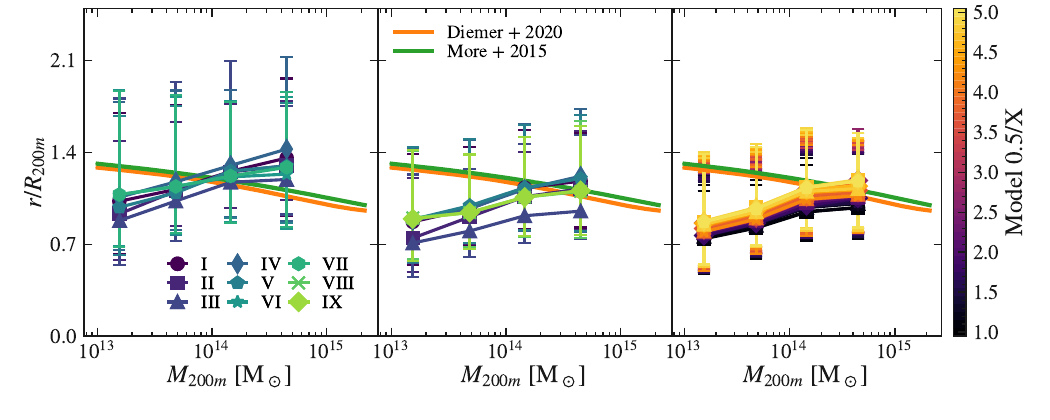}
    \caption{The leftmost and the middle panels both show the dependence of transition region on host mass derived using different feature combinations. The leftmost panel is generated by models using \textsc{SubLink} based labels, while the middle panel represents models using the fiducial distance-based label $0.5/5.0$. The x-axis denotes the mass of the cluster, and the y-axis the normalised distance. Each mass bin median zero-point and its 16th-84th percentile scatter is mapped to the median mass in that mass bin. Comparison is made with $R_{sp}$ reference curves from literature, computed using cluster $M_{200\rm m}$. The mass dependence of $r_0$ is opposite to that of both estimated splashback definitions. The error bars indicate intrinsic scattering. The rightmost panel shows the same mass dependence of $r_0$ from all distance-based models. The Roman numerals referenced across the leftmost and middle panels are as follows: I. Six properties, II. All properties, III. Dynamical properties, IV. Gas properties (without sSFR), V. Gas properties (with sSFR), VI. Stellar properties (without sSFR), VII. Stellar properties (with sSFR).}
    \label{fig:zeros-sublink_dist}
\end{figure*}

Once a galaxy enters the cluster environment, it experiences a series of quenching mechanisms exerted by the intracluster medium, one of them being ram pressure stripping. Therefore, understanding how cluster-driven processes act directly upon galaxy properties illuminates the spatial extent of the cluster environment's influence. In this section we quantify the radial reach of the ``cluster environment'' using two complementary diagnostics derived from our probabilistic classifier: (i) stacked median profiles of the nonsigmoid probability $z(r)$, such as Figure~\ref{fig:profiles-combined}, and (ii) the cluster-by-cluster zero-crossing radius $r_{0}$ compared with splashback radius definitions, as shown in Figure~\ref{fig:zeros-sublink_dist}. 

Distances are expressed as $r/R_{200,\rm mean}$, with $R_{200,\rm mean}$ defined for the host cluster of each galaxy. All host clusters are distributed across four mass bins. The results are produced by the fiducial distance based model denoted by $0.5/5.0$ and the $\textsc{SubLink}$ based model. For a given model and mass bin, we compute (1) the halo-level $z(r)$ profiles, binned in $\log r$, and (2) the median and 16-84th percentile envelopes across cluster mass bins. Figure~\ref{fig:profiles-combined} displays the median $z(r)$ curve of model $0.5/5.0$. This probability profile has the following morphology: an inner region where $z<0$, followed by a rapid rise between $\sim0.5$ and $\sim1.5\,R_{200,\mathrm{mean}}$ where the distribution passes $z = 0$, and an outer region beyond $\sim2\,R_{200,\mathrm{mean}}$ where $z$ asymptotically approaches positive values. The distributions of lower-mass hosts tend to show slightly higher $z$ at fixed radius inside $R_{200,\mathrm{mean}}$. Overall the mass bin distributions exhibit uniformity across different distance based models. 

Figure~\ref{fig:zeros-sublink_dist} summarises the dependence of $r_0$, the cluster-to-cluster zero points predicted by models trained on two labelling schemes as a function of the median mass in each cluster mass bin and the complementary convergence result. The \textsc{SubLink}-based model and the fiducial $0.5/5.0$ model are trained with the same seven feature combination. In this section, we discuss the transition region derived from models trained on the six primary properties (denoted by I).  We observe that $r_0$ increases from $\sim 0.8\,R_{200,\mathrm{mean}}$ in the lowest-mass bin to $\sim 1.0$--$1.1\,R_{200,\mathrm{mean}}$ at $\rm M_{200,\rm mean}\sim10^{14.5}\,M_\odot$, with a possible flattening at the highest masses. The vertical error bars (16--84th percentiles) are wide, spanning approximately $0.5$--$1.4\,R_{200,\mathrm{mean}}$, underscoring cluster-to-cluster diversity and intrinsic scattering of galaxy members within clusters at the location of the environmental transition. This intrinsic scatter is discussed in Section~\ref{sec:scatter}. The models are nearly indistinguishable at the level of their medians, emphasising that moderate changes in label construction do not qualitatively alter the inferred boundary at the upper end of our radial tests. All distance-to-mass trends hereafter are overlaid with the analytic splashback predictions from \citet{More2015} and \citet{Diemer2020}, which decline monotonically with halo mass. The ML-predicted $r_0$ resides in a similar radial domain but follows an opposite mass trend, indicating that the zero-crossing of $z(r)$ is \emph{not} a direct proxy for the caustic-defined splashback radius. Instead, it marks the radius where the classifier's learned representation for ``cluster-like'' vs. ``field-like'' galaxies balances, a quantity sensitive to both physical processes (e.g. environmental quenching, stripping), on top of measurement and selection effects encoded in the training set. We adopt $r_0$ as our fiducial scale for the property-based measure of the spatial extent of the cluster environment's influence on galaxies.  

These findings caution against treating $R_{200,\mathrm{mean}}$ as a hard environmental boundary in either simulations or observations. Instead, the transition appears extended and probabilistic, with a non-negligible fraction of environmentally affected galaxies beyond $R_{200,\mathrm{mean}}$. While splashback-based radii provide a physically motivated dynamical scale, our classifier-predicted $r_0$ traces the imprint of environment on galaxy properties, and the two need not coincide. In the splashback radius estimation scheme described in \citet{Diemer2020}, a broad transition is also inherent due to the requirement of choosing a percentage of orbits to enclose. In summary, the cluster environment influences our chosen list of galaxy properties out to $\sim 1 \text{--} 1.2 R_{200,\mathrm{mean}}$ on average, with scattering on cluster and galaxy levels. Both stacked profiles and zero crossings consistently support this picture, and the trends are robust across our best-performing models. This probabilistic view complements splashback measurements and demonstrates a data-driven way to probe the spatial extent of the environmental impact.

\subsection{Intrinsic scatter}\label{sec:scatter}
Now we interpret the vertical error bars shown in Figure~\ref{fig:zeros-sublink_dist}. Using logit bins, we can directly investigate how the network's prediction correlates with a galaxy's spatial distribution. Low logits ($\logit\ll0$) yield tightly clustered, inner–halo samples, while high logits ($\logit\gg0$) pick out galaxies in the field region. The intermediate logits around zero correspond to the broad, transition zone near the virial boundary at cluster outskirts, where samples represent the physical detection of a mixed population. 

\begin{figure}
    \centering
    \includegraphics[width=\linewidth]{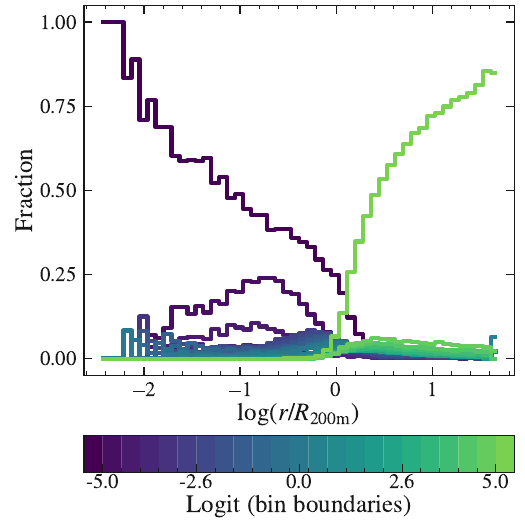}
    \caption{Radial distribution of galaxy logit output ranges from the fiducial model ($0.5/5.0$ with six properties). The curves show the normalised contribution of galaxies within a sequence of logit bins as a function of $\log(r/R_{200,\mathrm{mean}})$. The colour bar denotes the boundary of bins. There are $[-\infty, -5.0]$ on the far left end (purple) and $[5.0, \infty]$ on the far right end (light green). In between there are 21 logit bins with inner step $\approx 0.4$. The logit bin $[-0.23, 0.23]$ encloses the zero point, which is demarcated in the middle.
    Therefore, the dark purple curve indicates confidently cluster-like samples ($\mathrm{logit} \in [-\infty, -5.0]$),  and light green represents field-like samples ($\mathrm{logit} \in [5.00, \infty]$). In between, the logit bins show intrinsic scattering of galaxy samples, where $[-0.23, 0.23]$ contains the most ambiguous sample. This shows that there is intrinsic scattering of ambiguous galaxy samples across all radial bins, indicating that galaxies are impacted by the cluster environment through a gradual process.}
    \label{fig:intrinsic_scatter}
\end{figure}

Figure~\ref{fig:intrinsic_scatter} reveals three distinct regimes in the classifier’s notion of “cluster” versus “field” membership. The dark purple curve indicates confident cluster members. These are objects with very negative logits in $[-\infty, -5]$. Their normalised histogram (purple) peaks at $r/R_{200,\mathrm{mean}}\lesssim 1$ and then falls off sharply at larger distances. This confirms that when the network is highly confident of a cluster member, it indeed selects objects deep inside the virialised region. 

The light green curve delineates the normalised spatial distribution of confident field galaxies. These have very positive logits in $[5.0,\infty]$, and their distance distribution is nearly zero inside $1\,R_{200,\mathrm{mean}}$, then gradually rises beyond. 

The ``ambiguous population'' is represented by the sequence of non-overlapping logit bins created in between $[-5.0, 5.0]$. Galaxies with logits near zero in the range $[-0.2,0.2]$, represented by the turquoise green curve, form a low‐amplitude, broad distribution spanning both inside and outside $R_{200,\mathrm{mean}}$. These are precisely the galaxies that are still undergoing transition, as they enter a "preprocessing" zone in the cluster outskirts. It is clear that they are distributed across the cluster environment, therefore inducing the intrinsic scattering across the transition region. 

Taken together, all curves demonstrate that our network has learned a physically meaningful transition. The ``ambiguous'' region around $\logit\approx0$ shows that neither purely dynamical nor purely environmental indicators can unambiguously distinguish members from interlopers. This further reinforces the importance of establishing a probabilistic boundary for determining cluster memberships.

It is important to note that the dark purple line and the light green line intersect slightly beyond $R_{200,\rm mean}$. This indicates that this region contains a fraction of both confident cluster and field members, further suggesting that the galaxy membership transition likely takes place here. This suggests that the set of galaxy properties we feed into the network (stellar-to-subhalo mass ratio, gas fraction, colour, and metallicities...) exhibit gradual changes across the $R_{200,\rm mean}$. Each physical category of properties also experience change at different locations and under various time scales, due to the complexity in quenching mechanisms galaxies may experience. This is explained in detail in Section~\ref{sec:prop_based}.

The outliers such as backsplash galaxies are also successfully captured by the model's logit predictions. The classifier’s output can itself serve as a {\it data‐driven} definition of a halo boundary: selecting objects with $\logit<0$ recovers a sample confined within the “true” virial region, while $\logit>0$ picks out the infalling population.  
This shows that the model correctly identifies a physically distinct population whose properties do not exactly fit the "cluster" or "field" categories. The intrinsic scatter feature is inherent regardless of labels, as indicated by the convergence test (Section~\ref{sec:conver_test}). This ambiguity is an expected physical result, reflecting the known complexity of the cluster outskirts: a mixture of infalling and orbiting galaxy populations \citep[e.g.][]{Aung2020,Aung2023,Farid_2023} and the fact that the gas does not follow the dark matter in this region \citep{Aung2021}.

\subsection{Mass dependence of the transition region}\label{sec:mass_trend}
Now we discuss the the meaning of the mass trend displayed across all figures on our ML-learned radius-mass relation. We found that mass dependence of the transition region is related to the location ram pressure occurs for clusters of different masses and scales with mass concentration. 

According to existing theory, dark‐matter haloes exhibit a well‐established mass-concentration relation of the form $
c(M) \propto M^{-\alpha}$ with $\alpha \approx 0.1
$
\citep[e.g.,][etc.]{Bullock2001,Dolag2004,Dutton2014}. As haloes of different masses collapse at different epochs 
in hierarchical structure formation, lower-mass haloes typically form earlier, when the mean cosmic density is higher. This leads to more compact central regions, indicating higher concentrations. Their high-mass counterparts, however, tend to assemble later, producing shallower inner profiles. 
In the NFW formalism, this means that massive haloes have larger scale radius $r_s$, which implies that their density profiles decline more gradually with radius compared to low-mass haloes \citep{Navarro1996}. Consequently, lower‐mass haloes are on average more centrally concentrated than their higher‐mass counterparts. 

Since the network is trained on galaxy properties, the transition it learns should reflect when and where such properties respond to the cluster environment's impacts, one of which is ram-pressure stripping. In the classical ram‐pressure stripping framework established in \citet{Gunn1972}, a galaxy moving at velocity $v$ through an intracluster medium (ICM) of density $\rho_{\rm ICM}(r)$ will lose its gas when 
\begin{equation}
  \rho_{\rm ICM}(r)\,v^{2}
  \;\gtrsim\;
  2\pi G\,\Sigma_{\star}\,\Sigma_{\rm gas}\,,
\end{equation}
where $\Sigma_{\star}$ and $\Sigma_{\rm gas}$ are the stellar and gas surface densities of the galaxy. In a high‐concentration halo, we know that $\rho_{\rm ICM}(r)$ declines steeply with radius, so the condition for ram-pressure stripping and external pressure confinement is met only at small $r/R_{200}$.  By contrast, lower‐concentration halos possess shallower gas density profiles, making it confine gas out to larger fractions of $R_{200}$. Therefore, ram pressure also tends to strip further out. The extent of ram pressure stripping can also be considered using the relative velocity of galaxies in high-mass clusters compared with those that reside in low-mass cluster environments. The characteristic galaxy-ICM relative velocity is found to scale as $v \propto M^{1/3}$. This is confirmed in large $N$-body and hydrodynamic simulations for dark matter, subhalos, and galaxies \citep{Evrard2008,Munari2013} and in observations
\citep{Saro2013}. This relation has also been recently confirmed in IllustrisTNG \citep{Sohn2022}.

\begin{figure*}
    \centering
    \includegraphics[width=\linewidth]{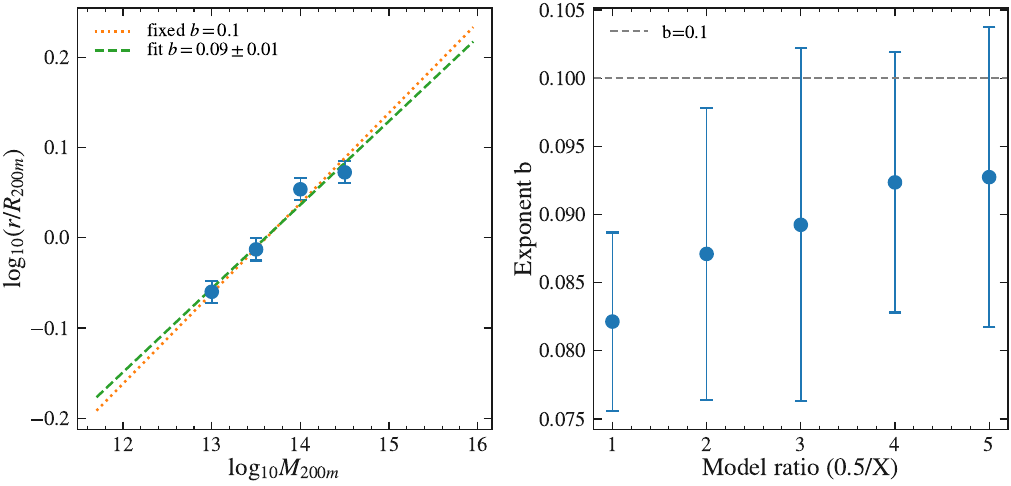}
    \caption{The left panel shows the logarithmic fit ($y = c+bx$) with errors on the transition region curve predicted by the fiducial $0.5/5.0$ model. The $x$-axis is $M_{200, \rm mean}$ in log-space, and the $y$-axis is the normalised distance in log-space. Points represent the bin medians with $1\sigma$ uncertainties. The orange line is produced by a fixed-fit using $b = 0.1$ (reduced $\chi^2 = 1.082$). The green line represents a free-fit to find the best $b$ (reduced $\chi^2 = 1.406$). The right panel documents the free-fit $b$ as a function of model ratio with error and $b = 0.1$ as the reference line. This shows that the mass trend of the transition region predicted by the proposed ML algorithm scales with $x^{0.1}$.}
    \label{fig:mass_fit}
\end{figure*}

In Figure~\ref{fig:mass_fit}, we show the best‐fit transition radius $r_0$ (where $\logit=0$) as a function of halo mass $M_{200,\rm mean}$.  We find:
\begin{equation}
  r_0\;\propto\;M_{200m}^{\;\alpha},\text{ such that } \alpha\approx0.10\,.
\end{equation}
This behaviour mirrors the well‐known decline of halo concentration with mass, where $c(M_{200m})\propto M_{200m}^{-\beta}$, and $\beta\approx0.1$ in both dark‐matter–only and hydrodynamical simulations 
\citep{Bullock2001, Maccio2008}.
Indeed, since more concentrated systems confine their hot gas more tightly, ram‐pressure stripping and related environmental processes can only act out to smaller radii \citep{Bahe2015}.  Lower‐mass (higher‐concentration) halos retain steep inner profiles, whereas higher‐mass (lower‐concentration) systems exhibit shallower outskirts that permit gas removal—and thus galaxy membership transition—to extend to larger fractions of the virial radius.

Quantitatively, if $c\propto M^{-\beta}$ then one expects
\begin{equation}
  r_0 \;\sim\;\frac{1}{c}\;\propto\;M^{\,\beta}\,
.
\end{equation}
This is in excellent agreement with our fit $\alpha\approx\beta\approx0.10$.  This correspondence demonstrates that the classifier’s learned boundary is not an arbitrary threshold but directly reflects the underlying gravitational potential and its mass‐dependent concentration.

\subsection{Relating to established boundary definitions}\label{sec:compare}
We now evaluate how well established and physically motivated definitions of the cluster boundary, namely the spherical overdensity radii $R_{200,\rm mean}$, $R_{200,\rm crit}$, and the dynamically motivated splashback radius $R_{\rm sp}$, distinguish cluster members from field galaxies when used as binary labelling thresholds. For each boundary definition, we assign galaxies lying within the radius definition cluster galaxy labels and those outside field galaxy labels (referred to herein as the $1/1$ labelling scheme).  We then train and evaluate models using all combinations of input features to see whether these boundary definitions yield clean separations in the learned transition region. By comparing the inferred zero‐crossing radii, classification accuracy, and probability‐profile sharpness against our network’s predictions, we assess whether $R_{200,\rm mean}$, $R_{200,\rm crit}$, or $R_{\rm sp}$ alone captures the true cluster–field split or whether more flexible, property-based boundaries are required.  

\begin{figure}
  \centering
  \includegraphics[width=\linewidth]{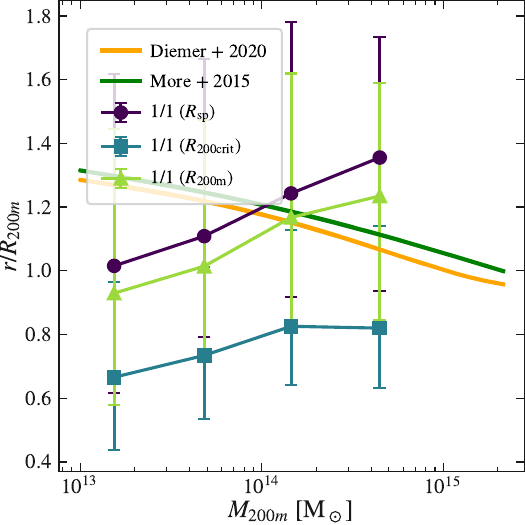}
  \caption{Dependence of the transition region on host mass using 1/1 labels based on three established cluster boundaries (\(R_{\mathrm{sp}}\), \(R_{200,\mathrm{mean}}\), and \(R_{200,\mathrm{crit}}\)) with six primary properties. This is overlaid with the same reference curves from \protect\citet{Diemer2020} and \protect\citet{More2015}. If these boundaries are good demarcations to determine cluster and field galaxies, then a 1-to-1 label scheme should result in a clean separation on this end, where the transition point for each mass bin equals to the value of the particular boundary of clusters with the same mass. There also should be no error bars. This indicates that these boundary definitions do not distinctly separate cluster and field galaxies based on their intrinsic properties.}
  \label{fig:1to1-sixprop}
\end{figure}

Figure~\ref{fig:1to1-sixprop} shows that the chosen boundaries do not explicitly separate cluster and field galaxies when classified using the six primary properties we use in the fiducial model. If the chosen boundaries distinctly separate cluster and field galaxy memberships, then the model learned transition should correspond to the original label scheme. This is due to the non-spherical nature of cluster boundaries, as well as the trajectory of galaxies as they enter a cluster environment. As indicated in \citet{O’Kane2024}, galaxies that reside in the filaments have different intrinsic properties (i.e. star formation rate) compared to those in the field, suggesting that preprocessing occurs beyond defined cluster boundaries. This indicates that the transition is not marked by a sharp boundary, but  a probabilistic one. Nonetheless, spherical overdensity based definitions (i.e. $R_{200, \rm mean}$ and $R_{200, \rm crit}$) capture the region where the cluster potential dominates and are relatively consistent across simulations and observations. The dynamically motivated splashback radius, through making a sharp drop in the halo density profile, encloses the outer boundary of recently accreted material. Therefore, these boundary definitions are able to serve as benchmarks when investigating the spatial extent of the cluster environment, even though they may not capture the changes in intrinsic properties that galaxies experience.

\subsection{Property‑based transition boundaries}\label{sec:prop_based}
Finally, we incorporate the insights gained from our feature-combination tests by investigating the transition boundary inferred when using all combinations of galaxy properties, organised by their underlying physics. In particular, we categorise input features into three classes—stellar properties, gas properties, and dynamical properties—and train separate models for each category to isolate the contribution of each physical component to the cluster–field galaxy transition. By comparing the resulting probability profiles and property-based transition regions, we assess whether stellar mass and metal content, gas mass and metal content, or dynamical indicators produce different transition regions.

\begin{figure*}
    \centering
    \includegraphics[width=\linewidth]{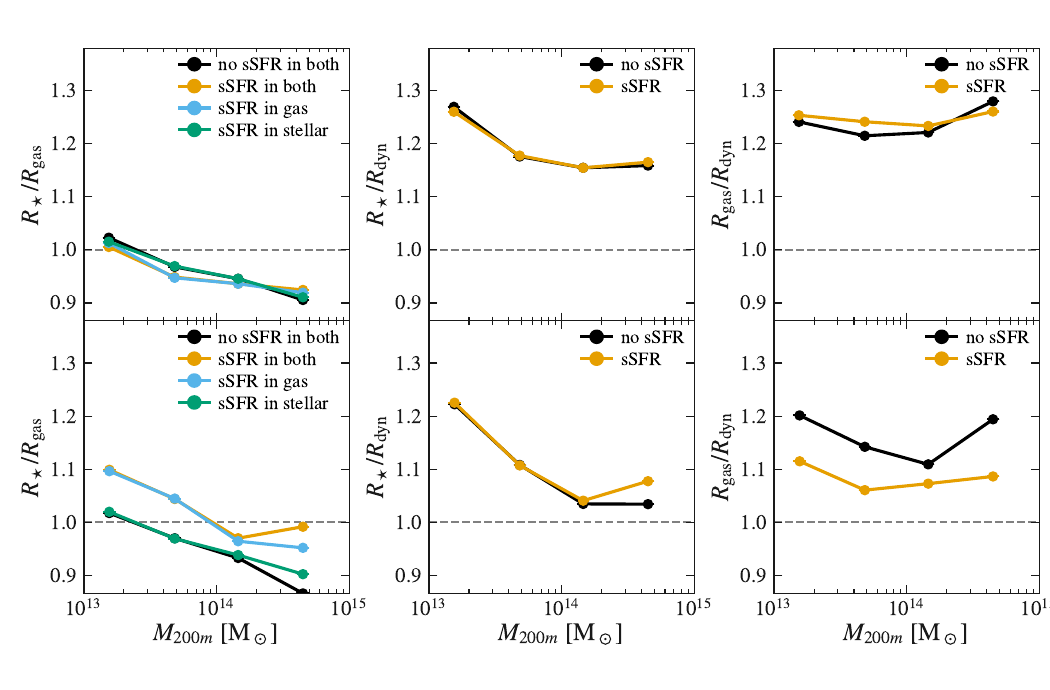}
    \caption{
\textbf{Median ratios of zero-crossing (transition) radii measured from different property categories.}
The top row uses the fiducial distance-based $0.5/5.0$ label, and the bottom row uses the \textsc{SubLink} label. Within each row, the columns correspond to different radius ratios, where \textbf{the leftmost column} is $R_\star/R_{\rm gas}$, \textbf{middle column} is $R_\star/R_{\rm dyn}$, and \textbf{the rightmost column} is $R_{\rm gas}/R_{\rm dyn}$. The variables $R_\star$, $R_{\rm gas}$, and $R_{\rm dyn}$ denote the transition radius inferred using the stellar-, gas-, and dynamical-intrinsic galaxy-property feature sets, respectively. The ratios are plotted as a function of cluster mass $M_{200,\mathrm{mean}}$. For each mass bin we take the \textbf{median ratio across clusters} (connected by lines). The horizontal dashed line marks when the the two property categories have identical transition radii. Because sSFR can be treated as either a stellar- or gas-related feature, we show several model variants as denoted by the legend labels. In the stellar/gas ratio panels, four curves indicate whether sSFR is excluded from both sets, included in both sets, or included in only one set. In the stellar/dynamical  and gas/dynamical panels, two curves indicate whether sSFR is included or excluded from the relevant baryonic property set. Overall, deviations from unity indicate offsets between the radii at which different property categories transition, consistent with different physical processes acting over different spatial and temporal scales.}
    \label{fig:compare_dist}
\end{figure*}

Figure~\ref{fig:zeros-sublink_dist} shows the transition regions predicted by different feature combinations (II - VII). To further visualise the offsets between the chosen physical processes, Figure~\ref{fig:compare_dist} demonstrates ratios computed among the $r_0$ of different feature combinations.

The ordering of these transition radii broadly follows the expectation from environmental‐process theory: dynamical indicators shift deepest in the potential well, while it is harder to disentangle where stellar and gas properties transition. The physical change in each component depends on the ram-pressure and strangulation galaxies experience as they enter the ICM, which can deplete or heat gas beyond $R_{200,\rm mean}$ \citep{Bahe2013}. In comparison, dynamical heating and orbital isotropisation occurs largely after pericentric passages \citep{Mamon2019,Oman2016, Oman2020}. When considering the manifestation of these changes, the sequence compresses inward: gas is depleted and stripped once a galaxy enters the ICM, then followed by change reflected by star‐formation‐based stellar indicators due to a quenching delay of $~2-4 $ $\rm Gyr$, and lastly morphological transformations. \citep[e.g.][]{Wetzel2013,Oman2016,Fossati2017}. However, our results show that the relative ordering of gas and stellar transition radii also depends on the mass of the cluster environment. For low-mass clusters, the gas-based transition region is at similar locations as the stellar-based one. Then, for clusters with higher masses, the expected sequence is more obvious. Ambiguity within such ordering also arises from whether the specific star formation rate (sSFR) is considered a gas or stellar property. In IllustrisTNG, SFR is calculated within the gas particles, so sSFR is computed as the ratio between a gas property (SFR) and a stellar property ($M_{\star}$) within the simulation. We therefore compare different feature combinations where sSFR could either be consider a stellar or gas property to disentangle the gas and stellar transitions. 

\section{Discussion of limitations and future work}\label{sec:discussion}
In this work, we present a data-driven framework to probe the transition region between cluster and field galaxies in galaxy clusters. The results are not designed to be interpreted as a hard cluster boundary, as we discover an intrinsic scattering. The results depend on the criteria used to classify galaxies as cluster-like versus field-like, something that is not universally or uniquely defined. Therefore, the result is slightly sensitive to how cluster and field galaxies are defined. 

The central question addressed here is: \emph{Can galaxy membership probabilistically inform the spatial extent of a cluster’s environmental influence?} Future work could be proposed to resolve the reciprocal problem: \emph{Can a boundary inferred purely from galaxy properties be used to determine cluster membership?} One possible direction is to employ unsupervised clustering techniques within the high‐dimensional galaxy property space to identify spatial separations that may correspond to physical boundaries. Such methods could yield a more objective, data‐driven spatial demarcation between cluster and field populations, enabling a direct comparison between spatially and property‐based definitions of cluster extent. It is important to note that the properties incorporated in this study are difficult to disentangle.

Beyond the mass-based cluster environment selection criteria done in this work, we can also choose cluster environments based on whether they host AGNs \citep{Wing2011, Galametz2012,Wylezalek2013,Croston2019,Golden-Marx2019}. This allows us to further probe black hole related quenching mechanisms and investigate the spatial effect of AGNs on galaxy intrinsic properties. 

This framework can also be applied to observations. During its observational adaptation, the training and validation sets of the DNN can be created with the help of cluster finders like the the redMaPPer \citep{Rykoff2014, Redmapper_Catalog}. The resulting probabilistic transition region can also be statistically weighted with the redMaPPer's outcome. Several observable properties can be chosen in this direction including multi-band colours, magnitudes, surface brightness, and spectral lines with photometric and redshift information. Projection effects (line-of-sight superpositions and photometric-redshift uncertainties) can lead to interloper contamination in galaxy membership assignments, which boosts the inferred richness and increases the scatter in any inferred transition radius. Therefore, the probabilistic transition region will be broadened due to potential biases compared to the intrinsic 3D case \citep{Rozo2015,Sunayama2020,Costanzi2019,Myles2025}.

Redshift evolution is also worth investigating since cluster environments evolve significantly with time \citep{ONeil2024}. Comparisons with other simulations can be conducted to reveal more information, especially by adding dwarf galaxies to illuminate stronger implications of the cluster environment's impact on low-mass systems. 
\section{Conclusions}
\label{sec:conclusions}
In this paper, we use the largest simulation volume of the IllustrisTNG simulation suite, TNG300-1, to probe the spatial extent of the cluster environment's impact on galaxy evolution. To overcome the resolution limit, we choose galaxies with $M_{\star} > 10^{9.5}\,M_{\odot}$ and $M_{dm} > 10^9\,M_{\odot}$ that reside in clusters with $M_{200,\rm{mean}} > 10^{13} \, M_{\odot}$. For each galaxy, we define the distance-to-closest cluster by calculating the periodic distance with respect to all clusters, and record the lowest one. We then examined galaxy properties as a function of this distance normalised by $R_{200,\rm mean}$. We employ a data-driven framework that classifies galaxy samples into cluster and field galaxies using intrinsic galaxy properties. With this classifier, we determine a probabilistic transition region for cluster environments of four different mass ranges. With the mass dependence of transition regions, our results are compared with established cluster boundary definitions. Our findings are summarised in the following sections:
\begin{itemize}
    \item In Figure~\ref{fig:profiles-combined}, we provide the first quantitative, probabilistic map of the transition region, which directly links galaxy–cluster distance to membership probability. In Figure~\ref{fig:zeros-sublink_dist}, we further extract the cluster-to-cluster zero point median and with its the 16th-84th percentile errors to highlight the transition region marked by the cluster environment's influence on galaxies. It follows an increasing trend with the cluster environment mass, which is consistent with the location of ram pressure stripping for clusters of different masses. Figure~\ref{fig:mass_fit} shows that the mass dependence follows the power law $x^{0.1}$. It is also intrinsically scattered, which means that the process of a galaxy's properties changing when entering the cluster environment is gradual. 
    \item In Figure~\ref{fig:1to1-sixprop}, we investigate the effectiveness of established cluster boundary definitions in separating cluster and field galaxies. The results indicate that there are galaxies that live across these boundaries with mixed memberships. Therefore, these radii do not explicitly create a hard boundary that defines a bipartition for the two galaxy populations.
    \item Through categorising galaxy properties by their underlying physics, i.e. gas, stellar, and dynamical, we train category-specific models to determine transition regions of different physical processes, thereby isolating the cluster environment’s effect. Figure~\ref{fig:compare_dist} demonstrates that the dynamical properties are impacted deepest inside the cluster centre compared with the stellar and gas transitions. However, we find that the offset between gas and stellar transition regions is ambiguous due to various phenomenon. 
\end{itemize}

We find that there exists a probabilistic relationship between a galaxy's distance to its closest cluster and its cluster membership, derived based on intrinsic galaxy properties. Therefore, a transition region denoting the spatial extent of a cluster environment's influence on galaxies. This is revealed by our model's discovery of a physically distinct population whose intrinsic galaxy properties do not fit a "cluster" or "field" template. The mass dependence of this transition region is related to the location of ram pressure stripping and other quenching mechanisms. It also scales inversely with the mass-concentration relation. Furthermore, by varying the input intrinsic properties by their categories, we can also illuminate the underlying physical processes.

The probabilistic nature of the transition boundary in cluster outskirts is a signature of mixed populations. A possible future test would be to utilise kinematic data from ongoings surveys such as DESI \citep{ DESIInstru, DESIVali, DESIEarly, DESIdata1} or upcoming surveys such as Roman \citep{WFIRSTSpergel2015}. By first separating galaxies into infalling and orbiting samples, one could apply our classifier to each population independently. This would directly test whether the intrinsic scattering is resolved and reveal if each population has its own, sharper transition profile.
 
Furthermore, another powerful test of the physical nature of our transition region is to compare its location to independent measurements of other cluster boundaries. For instance, comparing our galaxy-property-based boundary to the accretion shock radius, as measured by stacked Sunyaev-Zel'dovich (SZ) effect measurements \citep[e.g.,][]{Anbajagane_2022}, provides strong evidence that our classifier is primarily sensitive to gas-related processes like ram-pressure stripping. In parallel, our property-based boundaries could be compared with splashback radius measured via weak lensing or galaxy dynamics. If our boundary aligns more closely with the gas shock than the dark matter splashback radius, it would strongly demonstrate that our classifier is primarily sensitive to gas-related processes like ram-pressure stripping.

\section{Acknowledgements}
We thank the Editor and the Referee for their careful reading and constructive comments. We thank Daisuke Nagai, Bhuvnesh Jain, Tri Nguyen, Lina Necib, Frank van den Bosch, and Jeffrey Kenney for helpful discussions. C.H. was supported by the Pamela Daniels'59 Fellowship at Wellesley College, the Wellesley College Career Education Grant, and the MIT Undergraduate Research Opportunities Program (UROP).
S.O. is supported by the National Science Foundation under Grant No. AST-2307787.
Some of the computations were performed on the Engaging cluster supported by the Massachusetts Institute of Technology.

We made use of the following software for the analysis:
\begin{itemize}
	\item {\textsc{Python}}: \citet{vanRossum1995}
	\item {\textsc{Matplotlib}}: \citet{Hunter2007}
	\item {\textsc{SciPy}}: \citet{Virtanen2020}
	\item {\textsc{NumPy}}: \citet{Harris2020}
	\item {\textsc{Astropy}}: \citet{Astropy2013,Astropy2018}
\end{itemize}

\section{Data Availability}
The data is based on the IllustrisTNG simulations that are publicly available at \url{https://tng-project.org} \citep{Nelson2019}.
Reduced data are available upon request.

\bibliographystyle{mn2e}
\bibliography{bibliography}

\appendix
\section{Distance algorithm}\label{apx:distance_alg}
\begin{algorithm*}
  \small
  \caption{Assign Galaxies to Closest Clusters and Calculate the Corresponding Distance}
  \label{alg:galaxy_distances}
  \KwIn{Positions $Gal\_pos$, groups $Group\_pos$, cluster indices $Cluster\_index$}
  \KwOut{Distances $gal\_distance$, closest groups $gal\_closest\_group$}
  
  $N \gets |Gal\_pos|$\;
  Initialise $gal\_distance$ with $\infty$ for all galaxies\;
  Initialise $gal\_closest\_group$ with $-1$ for all galaxies\;
  
  $D \gets (\text{BoxSize} \times 1000) / H$ \tcp*{Box size in kpc, $H$ is Hubble parameter}
  
  \For{$i \in Cluster\_index$}{
    \For{galaxy $j$}{
      $dx \gets |Group\_pos[i] - Gal\_pos[j]|$\;
      $dx \gets \min(dx, D - dx)$ \tcp*{Periodic boundary condition}
      $dist \gets \sqrt{dx \cdot dx}$\;
      \If{$dist < gal\_distance[j]$}{
        $gal\_distance[j] \gets dist$\;
        $gal\_closest\_group[j] \gets i$\;
      }
    }
  }
\end{algorithm*}
\section{Arcsinh transform}\label{apx:arcsinh_transform}
We adjust both the location and scale of each feature before applying the arcsinh transformation. This operation allows for the specific transformation parameters to be set.

For most properties going through arcsinh transformation, the distribution is shifted so that the median maps to zero. The interquantile range between the $q_{\mathrm{lo}}$ and $q_{\mathrm{hi}}$ quantiles is rescaled to map to a fixed 
dynamic range. For gas fraction, however, which is bounded ($0 \leq x \leq 1$) and heavily
concentrated near the upper bound, we instead define a fixed reference value $x_{\mathrm{c}}=1$ that 
is mapped to zero, while the interquantile span controls the scaling. This procedure maps the transformed distribution to exhibit approximate symmetry, such that it is centered on zero with 
comparable spread on either side and without one dominant tail. A different quantile range is also adopted to accommodate the original data structure.

Hereafter, let $Q_{\alpha}(Y)$\footnote{Here $Q_{\alpha}$ denotes $\alpha$-th sample quantile of the feature distribution. 
In general, for a random variable $X$ with a cumulative distribution function (CDF) $F(x)$, the $\alpha$–th quantile 
is defined as 
\[
q_\alpha = \inf\{\,x \in \mathbb{R} : F(x) \geq \alpha \,\} ,
\]
so that a fraction $\alpha$ of the data lies below $q_\alpha$. In practice, we estimate these values from 
the empirical distribution of the sample. For example, the 0.10 and 0.90 quantiles correspond to 
the values below which 10\% and 90\% of the data fall, respectively. Using quantiles rather than 
the minimum or maximum ensures robustness against outliers while capturing the central bulk of 
the distribution.
} denote the arbitrary $\alpha$‐th quantile of the set $Y$, and let $\tau$\footnote{The parameter $\tau$ (the \emph{target half--width}) specifies how wide the ``linear regime'' of the 
arcsinh transformation should be in the transformed space.} be the half‐width of the approximately linear region in $\tilde y$\footnote{The linear-like region is inherent to the arcsinh function.}. The first transformation we applied is a general arcsinh scaler, where we centre the feature so that its median maps to $0$. We apply this scaler to the specific star formation rate, the gas half mass radius, and gas metallicity, because they exhibit long-tailed behaviour in linear space and have non-positive values, making them unfit for the logarithmic transformation. In this scaler, we set
\begin{equation}
  y' = y + b\, ,
  \quad
  b = -\,\mathrm{median}(y)\, .
\end{equation}
We then compute lower and upper quantiles of the shifted data with
the lower quantile range $\alpha_{\rm lo}=0.05$ and the upper quantile range $\alpha_{\rm hi}=0.95$ to define
\begin{align}
  q_{\rm lo} &= Q_{0.05}(y') \nonumber\\
  q_{\rm hi} &= Q_{0.95}(y') \nonumber\\
  \Delta &= q_{\rm hi} - q_{\rm lo}.
  \label{eq:q}
\end{align}
We choose
\begin{align}
  a &= \frac{2\,\tau}{\Delta} \\
  \tau &= 2.0
\end{align}
so that approximately 90\% of the data span $[-\,\tau,\tau]$ in the transformed space.  The final mapping is
\begin{equation}
  \tilde y = \operatorname{arcsinh}\bigl(a\,y + b\bigr).
\end{equation}
In the case of gas fraction, $x = M_{\rm gas}/M_{\rm baryon}$, the distribution is bounded (typically $0 \le x \le 1$). In logarithmic scale, the distribution concentrates below $\log(x) = 0$, where $x = 1$.  The general arcsinh scaler is not capable of disentangling this highly dense region. Therefore, we employ a specialised scaler for gas fraction that shifts $x$ by a chosen centre $c$ and then applies an arcsinh transform.
  
Let the original feature be
\begin{equation}
  x \;=\; \frac{M_{\rm gas}}{M_{\rm baryon}},
  \quad 0 \le x \le 1.
\end{equation}
We must shift and scale so that $x=1$ maps to zero, and then apply an arcsinh to make the distribution more uniform.  Define
\begin{equation}
  y \;=\; x - c,
  \quad\text{with}\; c = 1.0,
\end{equation}
so that $y=0$ corresponds to $x=1$.  Next, compute
\begin{align}
  q_{\rm lo} &= Q_{0.10}(y) \nonumber\\
  q_{\rm hi} &= Q_{0.90}(y) \nonumber\\
  \Delta &= q_{\rm hi} - q_{\rm lo}.
\end{align}
Choose a target half‐width $\tau = 5.0$\footnote{This value is chosen based on trial and error. Similar to the previous scaler, this denotes the interval in which the data in the upper quantile spans.} in the (approximately) linear regime, and set
\begin{align}
  a &= \frac{2\,\tau}{\Delta} \\
  b &= -\,a\,c.
\end{align}
The final transformed variable is
\begin{align}
  \tilde y 
  &= \operatorname{arcsinh}\!\bigl(a\,(y - c)\bigr)\nonumber\\
  &= \operatorname{arcsinh}\!\bigl(a\,y + b\bigr).
\end{align}

Here we choose $c = 1.0$ so that the $y = 1$ maps to $\tilde y=0$, and choose quantiles $\alpha_{\rm lo}=0.10$ and $\alpha_{\rm hi}=0.90$ with a target half‐width $\tau=5.0$.  This construction ensures that deviations around $y = 1$ are treated symmetrically in the approximately linear regime of the arcsinh, while extreme tails are smoothly compressed. These parametrisations ensure that the majority of each feature’s dynamic range lies in the nearly linear regime, while compressing extreme tails.

\section{Performance metrics}\label{apx:performance_metrics}
The performance metrics of seven retrained models with different feature sets are shown in Table~\ref{tab:perf_metrics}.
\begin{table}
  \centering
   \caption{Property study: AUROC and PR–AUC for different feature sets. Highest values achieved across label sets are documented.}
  \label{tab:perf_metrics}
  \begin{tabular}{@{}lcc@{}}
    \toprule
    \textbf{Feature set} & AUROC & PR–AUC \\
    \midrule
    Six primary properties                                        &   0.989    &    0.996   \\
    \addlinespace[3pt]
    All fifteen properties                                        &    0.995   &      0.998 \\
    \addlinespace[3pt]
    Gas–only (without sSFR)                                                      &  0.986     &   0.994    \\
    \addlinespace[3pt]
    Gas–only (with sSFR)                                                      &    0.986  &   0.995    \\
    \addlinespace[3pt]
    Stellar–only  (without sSFR)                                                 &   0.984    &   0.994    \\
    \addlinespace[3pt]
    Stellar–only   (with sSFR)                                                 &   0.985    &    0.994   \\
    \addlinespace[3pt]
    \addlinespace[3pt]
    Dynamics–only                                                 &    0.984   &  0.993     \\

    \bottomrule
  \end{tabular}
\end{table}

\label{lastpage}

\end{document}